\documentclass[prb,a4paper,twocolumn,showpacs,superscriptaddress]{revtex4}

\usepackage{amsmath}
\usepackage{amssymb}
\usepackage{bm}
\usepackage{graphicx}
\usepackage{color}
\newcommand{\bfsfG}{\mbox{\sffamily\bfseries{G}}}
\newcommand{\bfsfI}{\mbox{\sffamily\bfseries{I}}}
\newcommand{\sssection}[1]{{\par\it #1.---}}
\begin{document}

\title{Green function surface-integral method for nonlocal response of plasmonic nanowires in arbitrary dielectric environments}
\author{Wei Yan}
\affiliation{DTU Fotonik, Department of Photonics Engineering, Technical University of Denmark, DK-2800 Kongens Lyngby, Denmark}
\affiliation{Center for Nanostructured Graphene, Technical University of Denmark, DK-2800 Kgs. Lyngby, Denmark}

\author{N. Asger  Mortensen}
\affiliation{DTU Fotonik, Department of Photonics Engineering, Technical University of Denmark, DK-2800 Kongens Lyngby, Denmark}
\affiliation{Center for Nanostructured Graphene, Technical University of Denmark, DK-2800 Kgs. Lyngby, Denmark}

\author{Martijn Wubs}
\email[]{mwubs@fotonik.dtu.dk}
\affiliation{DTU Fotonik, Department of Photonics Engineering, Technical University of Denmark, DK-2800 Kongens Lyngby, Denmark}
\affiliation{Center for Nanostructured Graphene, Technical University of Denmark, DK-2800 Kgs. Lyngby, Denmark}

\date{\today}

\begin{abstract}
We develop a nonlocal-response generalization to the Green-function surface-integral method (GSIM), also known as the boundary-element method (BEM). This numerically light method can accurately describe the linear hydrodynamic nonlocal response of arbitrarily shaped plasmonic nanowires in arbitrary dielectric backgrounds. All previous general-purpose methods for nonlocal response are bulk methods. We also expand the possible geometries to which the usual local-response GSIM can be applied, by showing how to regularize singularities that occur in the surface integrals when the nanoparticles touch a dielectric substrate. The same regularization works for nonlocal response.
Furthermore, an effective theory is developed to explain the numerically observed nonlocal effects. The nonlocal frequency blueshift of a cylindrical nanowire in an inhomogeneous background generally increases as the nanowire radius and the longitudinal wavenumber become smaller, or when the effective background permittivity or the mode inhomogeneity increase. The inhomogeneity can be expressed in terms of an effective angular momentum of the surface-plasmon mode.
We compare local and nonlocal response of free-standing nanowires, and of nanowires close to and on top of planar dielectric substrates. Especially for the latter geometry, considerable differences in extinction cross sections are found for local as compared to nonlocal response, similar to what is found for plasmonic dimer structures.

\end{abstract}

\maketitle

\section{Introduction}

Plasmonic (metallic) structures support a surface-plasmon (SP) resonance, i.e, coherent free-electron oscillations at the structure boundary.\cite{Maier:2007} With the SP resonance, electric fields can be localized to the deep subwavelength scale, and accordingly be enhanced dramatically. This leads to numerous applications, including signal transfer in nanoscale photonic circuits, few-molecule bio-sensing and nonlinear phenomena.\cite{Maier:2007,Gramotnev:2010}

For individual plasmonic nanostructures of size larger than typically $10\,\rm nm$, it is accurate to describe metals with a local bulk refractive index,\cite{Maier:2007} as evidenced by numerous experiments. With recent progress in nanofabrication techniques, the sizes of individual plasmonic nanostructures can be controlled down to the deep nanoscale, below $10\,\rm nm$, and their relative distances even below a single nanometer.\cite{Nelayah:2007,Nicoletti:2011,Scholl:2012,Ciraci:2012b,Kern:2012,Savage:2012,Scholl:2013} This brings us into a regime where the foundation for the local bulk theory is challenged, since nonlocal response and the quantum wave nature of free electrons start to play a role.\cite{Scholl:2012,Ciraci:2012b,Kern:2012,Savage:2012,Scholl:2013,Teperik:2013} The nonlocal effects that we study here are a consequence of the fact that light interacts with moving charges, and manifest themselves only in nanoplasmonic structures.
We neglect quantum tunneling effects, which for example for plasmonic dimer structures come into play for separations less than half a nanometer.\cite{Zuloaga:2009} Nonlocal effects show up and dominate for larger separations,~\cite{Toscano:2012a,Dong:2012} and continue to be important when entering the quantum tunneling regime.~\cite{Dong:2012}

A direct and simplest generalization of the local theory is the hydrodynamic Drude model (HDM), describing besides the usual electromagnetic waves also longitudinal waves in the free-electron plasma.\cite{Bloch:1933,Ying:1974, Eguiluz:1976,Boardman:1982,Raza:2011} In the HDM, it is predicted that the nonlocal response blueshifts the resonance peak, modifies the field enhancement, gives rise to new resonances above the plasma frequency, and drives the second-harmonic generation of the plasmonic structure.\cite{Ruppin:1992,Ruppin:1973,Ruppin:2001,Fuchs:1987,FernandezDominguez:2012,Marier:2012,Raza:2011,Toscano:2012a,Raza:2012a,Toscano:2012b,Mochan:1987,Abajo:2008,David:2011,Sipe:1980,Ciraci:2012a}

For a few regularly shaped free-standing structures, such as a slab, cylinder, and sphere, the linearized hydrodynamic scattering problem can be solved analytically, for example using Mie theory or transformation optics techniques.\cite{Ruppin:1973,Mochan:1987,Fuchs:1987,Ruppin:1992,Ruppin:2001,Abajo:2008,David:2011,Raza:2011,Raza:2012a,FernandezDominguez:2012}
For realistic complex-shaped structures on substrates on the other hand, the hydrodynamic Drude response must be calculated numerically. Within the framework of the local-response bulk theory, the numerical simulations of the optical properties of plasmonic structures are mature since well-developed methods exist, such as the finite-difference time-domain method (FDTD),\cite{Yee:1966,Taflove:2005} finite-element method (FEM),\cite{Jin:2002} and the Green function surface-integral method (GSIM) which is also known as the boundary element method (BEM).\cite{Abajo:1998,Abajo:2002,Abajo:2006,Thomas:2007,Jung:2008,Kern:2009,Gallinet:2010} By contrast, few accurate numerical methods exist for the hydrodynamical response.

Recently, the FEM was generalized to calculate the hydrodynamic Drude response of arbitrary-shaped plasmonic structures.\cite{Toscano:2012a,Toscano:2012b,Ciraci:2012a,Hiremath:2012} The method was applied to nanowire dimers that show huge field enhancement,\cite{Toscano:2012a} to corrugated surfaces used for surface-enhanced Raman spectroscopy,\cite{Toscano:2012b} and to calculate extinction properties of V-grooves.\cite{Hiremath:2012} Very recently, the method was extended to calculate nonlocal effects in the waveguiding properties of plasmonic nanowires.\cite{Huang:2013,Toscano:2012c}

The FEM is a volume method, and nonlocal FEM in principle can handle both nanowires and three-dimensional structures. Yet it becomes numerically heavy for larger structures, especially for three-dimensional ones. This motivated us to develop a numerically lighter method. Already for local response it can sometimes be advantageous to turn to surface methods instead, where surfaces rather than scattering volumes need to be discretized. We started the present work anticipating that this advantage will only be greater for the nonlocal HDM, where a new length scale appears, namely the wavelength of the longitudinal waves. Since numerical meshes should be chosen considerably smaller than all length scales in the physical problem, in the HDM the meshing grid should be in the subwavelength scale of the longitudinal waves, which is below $1\,\rm nm$.~\cite{Toscano:2012a,Toscano:2012b} This suggests a larger relative advantage of surface methods for the hydrodynamic Drude theory.

Here we generalize the known Green function surface-integral (equation) method~\cite{Abajo:1998,Abajo:2002,Abajo:2006,Thomas:2007,Jung:2008,Kern:2009,Gallinet:2010} for local-response theories to include nonlocal response as described by the hydrodynamic Drude model. Moreover, we generalize the applicability of the usual local-response GSIM to an experimentally relevant class of geometries, namely where nanostructures rest on dielectric interfaces. These ``touching geometries'' may give rise to additional singularities in the surface integrals. We show how to regularize these singularities. The regularization procedure is the same in the local and nonlocal GSIM, and enables a convergent numerical implementation of the method.

Using our nonlocal GSIM, we investigate the effects of the nonlocal response on plasmonic nanowires, first for nanowires in a free-space background, and then for  nanowires above or resting on a dielectric substrate. In all our fully converged numerical calculations, the numerical grid size on the surface is in the subwavelength scale of the SP wave. We develop an approximate analytical theory for nonlocal blueshifts, and show its accuracy by comparison with our full GSIM numerics. We then use this theory to demonstrate how the strength of the nonlocal effects is determined by (i) the nanowire size $r_0$; (ii) the longitudinal wavenumber $k^{\rm L}$; (iii) the environmental permittivity $\epsilon_{\rm b}$ ; (iv) the angular momentum of the SP mode $l$.

The remaining part of the article is organized as follows.
Section~\ref{Sec:Nanowire_system} introduces the plasmonic nanowire structures under study and their environment.
In Section~\ref{Sec:HDM}, the hydrodynamic Drude model is introduced to describe the nonlocal response.
We generalize the GSIM to describe nonlocal response of nanowire structures in Section~\ref{sec:GSIM}.
The numerical implementation of the nonlocal GSIM is discussed in Sec.~\ref{Sec:numerical}, which also includes our new singularity regularization procedure that allows the GSIM to be applied to nanostructures that touch a dielectric interface.
In Section~\ref{sec:effective}, we develop approximate analytical expressions for nonlocal blueshifts for nanowires in inhomogeneous dielectric environments.
Testing the approximate theory is the red thread in our subsequent GSIM numerical simulations in Section~\ref{sec:numerical_results}, for nanowires without, above, and on dielectric substrates.
We summarize, conclude, and discuss our method and results in Section~\ref{Sec:discussion_and_conclusions}.
Some detailed derivations are relegated to Appendices~\ref{App:surface_integrals}-\ref{Sec:l_eff_def}.

\section{Nanowire System}\label{Sec:Nanowire_system}

\begin{figure}[h!]
\includegraphics[width=0.4\textwidth]{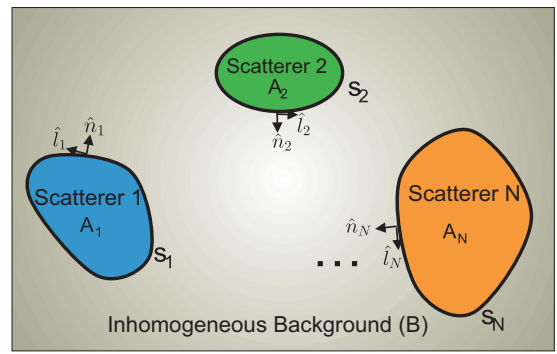}
\caption
{Illustration of the cross section of a nanowire system. $N$ isolated nanowires denoted as $A_{i}$ ($i=1,2,...N$) placed in an arbitrary inhomogeneous background denoted as B. The surface between the background and each nanowire scatterer is denoted as $S_i$. The $\hat n_i$ and $\hat l_i$ denote the unit vectors normal and tangential to $S_i$, respectively.}
\label{fig:1}
\end{figure}

We consider a nonmagnetic nanowire system, invariant in the $\hat z$ direction and with arbitrary cross section in the $\hat x, \hat y$ plane, see the sketch in Figure~\ref{fig:1}.
%
The system is divided into two regions: the plasmonic scatter region denoted as $A$, and the dielectric background denoted as $B$. Region $A$ consists of an arbitrary number of isolated plasmonic nanowires $A_i$ ($i=1,2,...N$). The individual nanowires are each a homogenous medium as described by the hydrodynamic Drude model. The dielectric function of the background is $\epsilon_b$, which we allow to be space dependent, and is assumed to be nonmagnetic. The boundary between $A_i$ and $B$ is called $S_i$. The outward-normal and tangential unit vectors at $S_i$ are denoted as $\hat n_i$ and $\hat l_i$ obeying $\hat n_i\times\hat l_i=\hat z$.

When exciting the system electromagnetically, for example with an electric current source $\mathbf J_b\exp(-i\omega t)$ in the region $B$, then the translation invariance suggests decomposing $\mathbf J_b$ into Fourier components along the wires,
\begin{eqnarray}
{\mathbf J_b}(\boldsymbol\rho ,z)=\int \mbox{d}{ k_z} { \widetilde{\mathbf J}_b}(\boldsymbol\rho ,{k_z})\exp \left( ik_zz \right),
\end{eqnarray}
where $\boldsymbol\rho$ represents $(x,y)$.
The interaction between $\mathbf J_b$ and the system is equivalent to a linear superposition of the sub-interactions between the $ \widetilde{\mathbf J}_b$ and the system. Each sub-interaction is a 2D problem in the $\hat x-\hat y$ plane with~\cite{Kong:2005}
\begin{subequations}
\begin{eqnarray}
\left(\boldsymbol\nabla_{\boldsymbol\rho}+ik_z\hat z\right)\times\mathbf E(\boldsymbol \rho)&=&i\omega\mu_0\mathbf H(\boldsymbol\rho),\\
\left(\boldsymbol\nabla_{\boldsymbol\rho}+ik_z\hat z\right)\times\mathbf H(\boldsymbol \rho)&=&-i\omega\mathbf D(\boldsymbol \rho)+\widetilde{\mathbf J}_b(\boldsymbol \rho,k_z),
\end{eqnarray}
\label{mastereq}
\end{subequations}
with $\boldsymbol\nabla_{\boldsymbol\rho}$ defined as $\hat x{\partial _x} + \hat y{\partial _y}$.

\section{Hydrodynamic Drude Model}\label{Sec:HDM}

 Plasmonic nanowires are of special interest owing to their ability to support SP resonances. We use the hydrodynamic Drude model (HDM) to describe the dynamics of the free electron gas.\cite{Bloch:1933,Ying:1974, Eguiluz:1976,Boardman:1982,Raza:2011,Ciraci:2012b} In the HDM, the electrons are collectively described by a density $n(\mathbf r,t)$ and velocity $\mathbf v(\mathbf r,t)$. The equation of motion is
\begin{equation}
{m_e}\left[ {\frac{{\partial \mathbf v}}{{\partial t}} + \mathbf v \cdot\boldsymbol \nabla\mathbf v} \right] = - \frac{{\boldsymbol\nabla {p_{\deg }}}}{n} + e\left( {\mathbf E + \mathbf v \times \mathbf B} \right),
\label{emotion}
\end{equation}
where $p_{\deg}$ is the pressure from the ground-state energy of the degenerate quantum Fermi gas, and we use $p_{\deg}$ in the Thomas--Fermi approximation. Using the charge conservation equation  $- {\partial n}/{\partial t} =\boldsymbol\nabla  \cdot n\mathbf v$, we linearize Eq.~(\ref{emotion}), and obtain the constitutive relation of the free-electron gas
\begin{eqnarray}
\frac{\beta^2}{\omega^2+i\omega\gamma}\boldsymbol\nabla\boldsymbol\nabla\cdot \mathbf P_f(\mathbf r)+\mathbf P_f(\mathbf r)=-\epsilon_0\frac{\omega_{p}^2}{\omega^2+i\omega\gamma}\mathbf E(\mathbf r),\label{hdm}
\end{eqnarray}
where $\omega_{p}$ represents the plasma frequency, $\gamma$ is the damping constant, and $\beta=\sqrt{3/5}v_{\rm F}$ with $v_{\rm F}$ the Fermi velocity. The operator $\boldsymbol\nabla\boldsymbol\nabla\cdot$ in Eq.~(\ref{hdm}) makes the  relation between the electric field and the polarization field a nonlocal one. Besides the free electrons, there are bound electrons, which constitutes another mechanism to polarize the metal with light. The constitutive relation of the bound electrons is
\begin{equation}
\mathbf P_d(\mathbf r)= \epsilon_0\chi_{\rm other}({\bf r},\omega)\mathbf E(\mathbf r),\label{lortz}
\end{equation}
a local relation, in contrast to Eq.~(\ref{hdm}).
The total polarization field $\mathbf P$ is $\mathbf P_d+\mathbf P_f$.

For infinite homogeneous systems (bulk metals), the polarization field  $\mathbf P$ can be uniquely decomposed into its transverse part $\mathbf P^{\rm T}$ with $\boldsymbol\nabla\cdot\mathbf P^{\rm T}=0$ and its longitudinal part $\mathbf P^{\rm L}$ with $\boldsymbol\nabla\times\mathbf P^{\rm L}=0$. By going to $k$-space, two independent solutions of the dispersion relations can be found, two types of waves corresponding to the transverse and the longitudinal dielectric functions
\begin{subequations}
\begin{eqnarray}
\epsilon _{\rm m}^{\rm T}(\omega) &=& \epsilon_{\rm other}(\omega)-\frac{\omega _{p}^2}{\omega ^2+ i\omega\gamma},\label{traep}\\
\epsilon_{\rm m}^{\rm L}(\omega) &=&\epsilon_{\rm other}(\omega)-\frac{\omega_{p}^2}{\omega^2+i\omega\gamma-\beta^2 k^2},
\end{eqnarray}
\label{epsilons}
\end{subequations}
with $\epsilon_{\rm other}=1+\chi_{\rm other}$. The dispersion of the transverse waves is $k(\omega)=\omega\sqrt{\epsilon_{\rm m}^{\rm T}(\omega)}/c$, while the dispersion of the longitudinal waves is determined by $\epsilon_{\rm m}^{\rm L}(\omega,k)=0$. Since they are independent solutions, the two types of waves do not interact with each other in infinitely extended metals.

Both types of waves also exist in finite homogeneous plasmonic structures, where they also propagate independently, except at boundaries.  Boundary conditions dictate the generation of mixed excitations: external light, a transverse wave, not only excites transverse but also longitudinal waves in the metal.\cite{Raza:2011,Barton:1979} The transfer-matrix method for nonlocal response of metal-dielectric multilayer structures illustrates this point quite well.\cite{Mochan:1987,Yan:2012} In our Green-function method below, we will also make use of this crucial fact that the transverse and longitudinal waves propagate independently within the homogeneous metal, but are not generated independently and at boundaries must occur in the right mixture so as to satisfy the boundary conditions.

\section{Green Function Surface Integrals}\label{sec:GSIM}

\subsection{Surface integrals for local response}\label{sec:GSIM_local}

We first give the known surface integrals for the local-response theory,\cite{Abajo:2002,Jung:2008,Kern:2009,Gallinet:2010} before introducing in Sec.~\ref{sec:GSIMnonlocal} the surface integrals for nonlocal response. In Ref.~\onlinecite{Jung:2008} it was stressed and shown that one of the advantages of the GSIM is that backgrounds such as infinite substrates can be taken into account in terms of their Green functions. The surface integrals here are valid for arbitrary spatially inhomogeneous backgrounds. We also allow light propagation in the direction along the nanowires ($k_z \ne 0$), thereby generalizing the results of Ref.~\onlinecite{Jung:2008} where light propagation in more than two dimensions is not considered. Additionally, we allow the possibility that the inhomogeneous background responds nonlocally, as discussed in Sec.~\ref{sec:GSIMnonlocal}. Finally, it was not clear until now how to apply the GSIM to nanostructures that touch a substrate. We solve  the associated mathematical difficulties in Sec.~\ref{Sec:numerical}, which enables us to present converged numerical results of the GSIM for touching geometries in Sec.~\ref{sec:numerical_results}.

\sssection{Surface integrals inside nanowires} First we give the surface-integral equations for the metal wires with the cross section $A_i$ and the boundary $S_i$. In the local description, the transverse fields $\mathbf E_i^{\rm T}$ and $\mathbf H_i^{\rm T}$ are coupled and the constitutive relation reads $\mathbf D_i^{\rm T}=\epsilon_0\epsilon_m^{\rm T}\mathbf E_i^{\rm T}$. The $\hat x,\hat y$-components of the fields can be expressed in terms of their $\hat z$-components.\cite{Kong:2005}
For positions $\boldsymbol \rho \in A_{i}$, the field components $E_{zi}^{\rm T}$ and $H_{zi}^{\rm T}$ satisfy the scalar surface integrals
\begin{subequations}
\begin{eqnarray}
E_{zi}^{\rm T}(\boldsymbol \rho)&=&-\oint_{{S_{i}}} \mbox{d}\boldsymbol{\rho'}\left[\,e_{i}^0(\boldsymbol \rho,\boldsymbol \rho') E_{zi}^{\rm T}(\boldsymbol \rho')+e_{i}^1(\boldsymbol \rho,\boldsymbol \rho'){E_{zi,n}^{\rm T}}(\boldsymbol \rho')\right],\nonumber\\
\label{trasieq1}\\
H_{zi}^{\rm T}(\boldsymbol \rho)&=&-\oint_{{S_{i}}}\mbox{d}\boldsymbol{\rho'}\left[\,m_{i}^0(\boldsymbol \rho,\boldsymbol \rho')  H_{zi}^{\rm T}(\boldsymbol \rho')+m_{i}^1(\boldsymbol \rho,\boldsymbol \rho') {H_{zi,n}^{\rm T}}(\boldsymbol \rho')\right],\nonumber\\
\label{trasieq2}
\end{eqnarray}
\label{trasieq0}
\end{subequations}
with the integration kernels
\begin{subequations}
\begin{eqnarray}
e_{i}^0(\boldsymbol \rho,\boldsymbol \rho') &=&m_{i}^0(\boldsymbol \rho,\boldsymbol \rho') =\hat n_{i}(\boldsymbol\rho') \cdot \boldsymbol\nabla_{\boldsymbol\rho'} {g_{i}^{\rm T}}(\boldsymbol \rho,\boldsymbol \rho'),\\
e_{i}^1(\boldsymbol \rho,\boldsymbol \rho') &=&m_{i}^1(\boldsymbol \rho,\boldsymbol \rho') =-{g_{i}^{\rm T}}(\boldsymbol \rho,\boldsymbol \rho').
\end{eqnarray}
\label{scacoeft}
\end{subequations}
Here, the scalar Green function $g_{i}^{\rm T}(\boldsymbol\rho,\boldsymbol\rho')$ satisfies $[\nabla_{\boldsymbol\rho}^2+(k_{\rho i}^{\rm T})^2]g_{i}^{\rm T}(\boldsymbol \rho,\boldsymbol \rho')=-\delta(\boldsymbol \rho-\boldsymbol \rho')$ and has the solution ${iH_0^{(1)}(k_{\rho i}^{\rm T}|\boldsymbol{\rho}-\boldsymbol{\rho}'|)}/{4}$ with $H_0^{(1)}$ being the zeroth-order Hankel function of the first kind; the subscript `$n$' in $E_{zi,n}^{\rm T}$ and $H_{zi,n}^{\rm T}$ stands for the directional derivative normal to the surface, e.g.,
$E_{zi,n}^{\rm T}=\hat n \cdot\boldsymbol\nabla_{\boldsymbol\rho} E_{zi}^{\rm T}$. The derivation of the surface integrals~(\ref{trasieq0}) is given in Appendix~\ref{sec:GSIM_local_eqs_metal}.

\sssection{Surface integrals outside of nanowires} Having discussed the surface integrals for the metal wires, we now turn to the background, which we allow to have an arbitrary spatially varying dielectric function $\epsilon_{b}({\boldsymbol \rho},\omega)$. This inhomogeneity makes the surface integrals more complicated than for the nanowires that we assumed homogeneous. For example, instead of scalar Green functions the surface integrals will feature tensor components of dyadic Green functions.  As derived in Appendix~\ref{sec:GSIM_local_eqs_background}, the surface integrals for the $\hat z$-components of the electric and magnetic fields are
\begin{widetext}
\begin{subequations}
\begin{eqnarray}
{E_{zb}}(\boldsymbol\rho )&=&E_{zb}^{\rm inc}(\boldsymbol\rho )+\oint\limits_S \mbox{d}\boldsymbol{\rho'} \,\left[{e_{b}^0(\boldsymbol\rho ,\boldsymbol\rho ')}{E_{zb}}(\boldsymbol\rho ') + {e_{b}^1(\boldsymbol\rho ,\boldsymbol\rho ')}{E_{zb,n}}(\boldsymbol\rho ')\right]
+\oint\limits_S \mbox{d}\boldsymbol{\rho'} \,\left[{f_{b}^0(\boldsymbol\rho ,\boldsymbol\rho ')}{H_{zb}}(\boldsymbol\rho ') + {f_{b}^1(\boldsymbol\rho ,\boldsymbol\rho ')}{H_{zb,n}}(\boldsymbol\rho ')\right],\nonumber\\
\label{trabaeq1}\\
{H_{zb}}(\boldsymbol\rho )&=&H_{zb}^{\rm inc}(\boldsymbol\rho )+\oint\limits_S \mbox{d}\boldsymbol{\rho'} \,\left[{m_{b}^0(\boldsymbol\rho ,\boldsymbol\rho ')}{H_{zb}}(\boldsymbol\rho ') + {m_{b}^1(\boldsymbol\rho ,\boldsymbol\rho ')}{H_{zb,n}}(\boldsymbol\rho ')\right]
+\oint\limits_S \mbox{d}\boldsymbol{\rho'} \,\left[{h_{b}^0(\boldsymbol\rho ,\boldsymbol\rho ')}{E_{zb}}(\boldsymbol\rho ') + {h_{b}^1(\boldsymbol\rho ,\boldsymbol\rho ')}{E_{zb,n}}(\boldsymbol\rho ')\right].\nonumber\\
\label{trabaeq2}
\end{eqnarray}
\label{trabaeq}
\end{subequations}
\end{widetext}
The $E_{zb}^{\rm inc}$ and $H_{zb}^{\rm inc}$ represent the $\hat z$-components of the incident electric and magnetic fields. Note that the integrations in Eq.~(\ref{trabaeq}) are over all metal-dielectric surfaces with $S= \sum_i S_{i}$. Again we wrote the integration kernels in short-hand notation. They are scalar functions, given in terms of components of the background dyadic electric and magnetic Green functions $\bfsfG _e$ and $\bfsfG_m$  (defined in Appendix~\ref{sec:GSIM_local_eqs_background}) and their spatial derivatives, i.e.,
\begin{widetext}
\begin{subequations}
\begin{eqnarray}
{(e,m)_{b}^0}(\boldsymbol\rho,\boldsymbol\rho') & = & \left[{\left( {i{k_z}\hat z - \boldsymbol\nabla _{\boldsymbol\rho'}} \right) \times \bfsfG _{e,m}^t}(\boldsymbol\rho,\boldsymbol\rho')\right]_{lz} +\frac{{i{k_z}}}{{k_{\rho b}(\boldsymbol\rho')^2}} \left[\hat l(\boldsymbol\rho') \cdot {\boldsymbol\nabla _{\boldsymbol\rho'}}{\left[ {\left( {i{k_z}\hat z - \boldsymbol\nabla _{\boldsymbol\rho'}} \right) \times \bfsfG _{e,m}^t}(\boldsymbol\rho,\boldsymbol\rho') \right]}\right]_{zz},\nonumber\\
 \label{bcoef1_a}\\
{(e,m)_{b}^1}(\boldsymbol\rho,\boldsymbol\rho') & = & -\frac{k_b(\boldsymbol\rho')^2}{k_{\rho b }(\boldsymbol\rho')^2} \left[{\bfsfG}_{e,m}(\boldsymbol\rho,\boldsymbol\rho')\right]_{zz}, \label{bcoef1_b} \\
{f_{b}^0}(\boldsymbol\rho,\boldsymbol\rho') & = & - i\omega {\mu _0} \left[{{\bfsfG }_e}(\boldsymbol\rho,\boldsymbol\rho') \right]_{zl}  + \frac{{\omega {\mu _0}{k_z}}}{{k_{\rho b}(\boldsymbol\rho')^2}}\left[\hat l(\boldsymbol\rho')\cdot {\boldsymbol\nabla _{\boldsymbol\rho'}}{{\bfsfG }_e(\boldsymbol\rho,\boldsymbol\rho')}\right]_{zz}, \label{bcoef1_c}\\
{f_{b}^1}(\boldsymbol\rho,\boldsymbol\rho') & = & \frac{{i\omega {\mu _0}}}{{k_{\rho b}(\boldsymbol\rho')^2}}\left[{\left( {i{k_z}\hat z - \boldsymbol\nabla _{\boldsymbol\rho'}} \right) \times \bfsfG _e^t}(\boldsymbol\rho,\boldsymbol\rho') \right]_{zz}, \label{bcoef1_d}\\
{h_{b}^0}(\boldsymbol\rho,\boldsymbol\rho') & = & i\omega {\epsilon _b}(\boldsymbol\rho')\left[{{\bfsfG }_m}(\boldsymbol\rho,\boldsymbol\rho') \right]_{zl}   -\frac{{\omega {\epsilon _b(\boldsymbol\rho')}{k_z}}}{{k_{\rho b}(\boldsymbol\rho')^2}}\left[\hat l(\boldsymbol\rho')\cdot {\boldsymbol\nabla _{\rho'}}{{\bfsfG }_m}(\boldsymbol\rho,\boldsymbol\rho') \right]_{zz}, \label{bcoef1_e}\\
{h_{b}^1}(\boldsymbol\rho,\boldsymbol\rho') & =& -\frac{{i\omega {\epsilon _b}(\boldsymbol\rho')}}{{k_{\rho b}(\boldsymbol\rho')^2}}\left[{\left( {i{k_z}\hat z - \boldsymbol\nabla _{\boldsymbol\rho'}} \right) \times \bfsfG _m ^t}(\boldsymbol\rho,\boldsymbol\rho')\right]_{zz}, \label{bcoef1_f}
\end{eqnarray}
\label{bcoef1}
\end{subequations}
\end{widetext}
where $k_{b}=\omega\sqrt{\epsilon_b}/c$ and $k_{\rho b}^2=k_b^2-k_z^2$. The superscript ``$t$'' in $\bfsfG ^t$ represents the transpose operation. The $\left[\bfsfG\right]_{lz}$ is the tensor component $\left[\bfsfG\right]_{lz}=\hat l(\boldsymbol\rho')\cdot\bigl\{\bfsfG(\boldsymbol\rho,\boldsymbol\rho') \bigl\} \cdot \hat z$, and  $\left[\bfsfG\right]_{zl}$ is analogously defined by $\left[\bfsfG\right]_{zl}=\hat z\cdot\bigl\{\bfsfG(\boldsymbol\rho,\boldsymbol\rho') \bigl\} \cdot \hat l(\boldsymbol\rho')$.

In the special case of a spatially homogenous dielectric background, the surface-integrals for the background become similar to those for the homogenous plasmonic scatterer in Eq.~(\ref{trabaeq}). In particular, the coefficients $f_b^{0,1}$ and $h_b^{0,1}$ vanish, while $e_b^{0,1}$ and $m_b^{0,1}$ assume the same forms as $e_i^{0,1}$ and $m_i^{0,1}$ in Eq.~(\ref{scacoeft}), just with $g_i^{\rm T}$ replaced by the background scalar Green function $g_b= {i}H_0^{(1)}({k_{\rho b}}|\boldsymbol\rho  - \boldsymbol\rho '|)/4$.

Returning to the general case of inhomogeneous dielectric backgrounds, one can split the dyadic Green function into ${\bfsfG }_{e,m}= {\bfsfG }_{e,m}^0+{\bfsfG }_{e,m}^s$, where ${\bfsfG }_{e,m}^0$ represents the dyadic Green function for a homogeneous background, and ${\bfsfG }_{e,m}^s$ represents the scattering contribution owing to the inhomogeneity in the background.\cite{Tomas:1995} The non-vanishing scattering contribution ${\bfsfG }_{e,m}^s$ gives rise to nonzero values for $f_b^{0,1}$ and $h_b^{0,1}$, and makes the other kernels more complicated. This is illustrated in Appendix B for the experimentally important example of a dielectric background consisting of a dielectric slab in air, i.e. a substrate layer that can support the plasmonic nanowires.

\sssection{Summary of local-response GSIM} We have now in Eq.~(\ref{trasieq0}) described the fields inside the metal wires as the surface integrals over the fields on the interior of their surfaces, and similarly Eq.~(\ref{trabaeq}) gives the fields in the dielectric background in terms of the fields on the surface exterior to these metallic nanowires. For a unique solution of the fields in all of space we need to specify boundary conditions that relate the fields on both sides of the interfaces. In the local-response approximation that we consider in this subsection, these are just the usual Maxwell boundary conditions, namely  that the tangential electric and magnetic fields be continuous across the boundaries.  We stress that in the above we arrived at a powerful generalization of the existing local-response GSIM,  by allowing the background dielectric function $\epsilon_{b}({\bf r})$ to have an arbitrary spatial dependence. The procedure is now to first solve for the fields on the surfaces, and after that to use these solutions in combination with the surface integrals to uniquely determine the fields in all of space. When solving for the fields on the surfaces, singularities in the integration kernels need to be dealt with. This is detailed in Sec.~\ref{Sec:numerical}, where it is also shown how to regularize additional singularities in case the surface touches a dielectric interface.

\subsection{Surface integrals for nonlocal response}\label{sec:GSIMnonlocal}

We now turn to the nonlocal-response theory and its associated surface integrals and boundary conditions. As was mentioned in Sec.~\ref{Sec:HDM}, in the hydrodynamic Drude model additional longitudinal waves exist in the metal, besides the usual transverse waves. These longitudinal and transverse waves propagate independently in the homogeneous metallic nanowires, except at their boundaries. The key insight leading to our Green function surface-integral method for the hydrodynamic model is then that for these longitudinal waves an additional surface integral can be formulated, independent of the other two, as presented below.

\sssection{Additional surface integral}
The longitudinal field $\mathbf E_{i}^{\rm L}$ by definition is rotation-free and in the plasmonic nanowire $A_i$ can thus be expressed in terms of a potential $\phi_i$ via
\begin{equation}
\mathbf E_{i}^{\rm L}=-\left(\boldsymbol\nabla_{\boldsymbol\rho}+ik_z\hat z\right)\phi_i,
 \end{equation}
where $\phi_i$ satisfies the scalar wave equation
\begin{equation}
\left(\nabla_{\boldsymbol\rho}^2+{k_{\rho i}^{\rm L}}^2\right)\phi_{i}(\boldsymbol \rho)=0,
\label{logweq}
\end{equation}
with ${k_{\rho i}^{\rm L}}^2={k_i^{\rm L}}^2-k_z^2$ and $\beta k_{i}^{\rm L}=({\omega ^2} + i\omega \gamma  - \omega _p^2/\epsilon_{\infty})^{1/2}$. The scalar Green function $g_{i}^{\rm L}$ associated with Eq.~(\ref{logweq}) is defined as the solution of
\begin{equation}
(\nabla_{\boldsymbol\rho}^2+{k_{\rho i}^{\rm L}}^2)g_{i}^{\rm L}(\boldsymbol \rho,\boldsymbol \rho')=-\delta(\boldsymbol \rho-\boldsymbol \rho').
\end{equation}
Directly analogous to the Green functions that we introduced before, the solution is given by
\begin{equation}
g_{i}^{\rm L}(\boldsymbol \rho,\boldsymbol \rho')=\frac{i}{4}H_0^{(1)}(k_{\rho i}^{\rm L}|\boldsymbol{\rho}-\boldsymbol{\rho'}|).
\label{loggeq}
\end{equation}
The main physical difference is that the longitudinal wavevectors $k_{\rho i}^{\rm L}$ are typically much larger than the transverse wavevectors $k_{\rho i}^{\rm T}$ of the metal and $k_{\rho b}$ of the dielectric background.
Analogous to the derivation of Eq.~(\ref{trasieq0}) in Appendix~\ref{sec:GSIM_local_eqs_metal}, we can now derive that the potential in the interior of the metal can be expressed as an integral over the same potential at the surface,
\begin{equation}
\phi_{i}(\boldsymbol \rho)=-\oint_{{S_{i}}} \mbox{d}\boldsymbol{\rho'}\,\left[ p_{i}^0(\boldsymbol\rho,\boldsymbol\rho')\phi_{i}(\boldsymbol \rho') + p_{i}^1(\boldsymbol\rho,\boldsymbol\rho') \phi_{i,n}(\boldsymbol \rho')\right],
\label{logsieq}
\end{equation}
with integration kernels
\begin{subequations}
\begin{eqnarray}
p_{i}^0(\boldsymbol\rho,\boldsymbol\rho')&=&\hat n_{i}(\boldsymbol\rho')\cdot \boldsymbol\nabla_{\boldsymbol\rho'} {g_{i}^{\rm L}}(\boldsymbol \rho,\boldsymbol \rho'),\\
p_{i}^1(\boldsymbol\rho,\boldsymbol\rho')&=&-{g_{i}^{\rm L}}(\boldsymbol \rho,\boldsymbol \rho').
\end{eqnarray}
\end{subequations}
Equation~(\ref{logsieq}) is the sought surface integral for the longitudinal fields in the plasmonic nanowire. The surface integrals Eq.~(\ref{trasieq0}) for the transverse fields in the metal and Eq.~(\ref{trabaeq}) for the fields in the background simply stay the same in the hydrodynamic Drude model.
Only if one would also wish to allow metal constituents also in the background, for example to describe an infinite metal substrate,\cite{Ciraci:2012b} and take its nonlocal response into account, would a modification  be needed for the background. We briefly discuss such a modification in Appendix~\ref{sec:GSIM_local_eqs_background}.

\sssection{Additional boundary condition}
Besides the three surface integrals~(\ref{trasieq0}), (\ref{trabaeq}), and (\ref{logsieq}), we again need boundary conditions to obtain unique solutions for the electromagnetic fields in all of space. In local-response theory we only needed the usual Maxwell boundary conditions, as we discussed in Sec.~\ref{sec:GSIM_local}, but for nonlocal response, additional boundary conditions (ABCs) are needed. In the present paper, we only consider metal-dielectric, not metal-metal interfaces. We also assume that the static free-electron density is a step function at the interface and constant within the metal, thereby neglecting Friedel oscillations and the electron spill-out associated with quantum tunneling on the sub-nanometer scale. These assumptions entail that only one ABC is needed for the hydrodynamic Drude model, which is the condition that the normal component of the free-electron current is continuous and hence by charge conservation vanishes at the boundary.\cite{Raza:2011,Jewsbury:1981a,Yan:2012} This condition can be combined with the usual Maxwell boundary condition that in the absence of free charges the normal component of the electric displacement field is continuous across the boundary, whereby the ABC can be unambiguously expressed as\cite{Yan:2012}
\begin{equation}
\epsilon_b\, \hat n\cdot\mathbf{E}_b= \epsilon_{\rm other}\,\hat n\cdot\mathbf{E}_i,
\label{ABCintermsofE}
\end{equation}
with $\epsilon_{\rm other}$ the bound-electron response of the metal as introduced in Eq.~(\ref{epsilons}). This ABC~(\ref{ABCintermsofE}) implies that in general the normal components of the electric field  makes a jump at the interface. Such a jump is the common situation also in the usual local-response approximation, but here in Eq.~(\ref{ABCintermsofE}) the jump is different than for local response where $\epsilon_{\rm other}$ on the right-hand side would be replaced by the full dielectric response $\epsilon _{\rm m}^{\rm T}$ of the metal, see Eq.~(\ref{traep}), including the Drude part for the free electrons.

In the ABC~(\ref{ABCintermsofE}), the electric field $\mathbf E_i$ at the interior of the metal interface of the $i^{\rm th}$ nanowire is the sum of the usual divergence-free electric field $\mathbf E_{i}^{\rm T}$ described by the surface integral~(\ref{trasieq0}) and of the (specifically hydrodynamic) rotation-free electric field $\mathbf E_{i}^{\rm L}$, described by the surface integral~(\ref{logsieq}) for its corresponding potential. The occurrence of this sum of independent solutions in a bounded region of space makes it intuitively clear that the ABC is needed for a unique solution in all of space.

\sssection{Summary of hydrodynamic GSIM}
In summary, three independent surface integrals~(\ref{trasieq0}), (\ref{trabaeq}), and (\ref{logsieq}) are needed for the hydrodynamic model, instead of the common first two for local response. These three integrals give rise to a unique and physically meaningful solution of the electromagnetic fields, when used in combination with three boundary conditions, two of which are the usual ones derived from Maxwell's equations. The third one is the additional boundary condition~(\ref{ABCintermsofE}), which is derived from local conservation of free charges after neglecting quantum spill-out of free electrons.


\section{Numerical implementation of nonlocal GSIM}\label{Sec:numerical}
For clarity, we first collect the surface integrals and boundary conditions needed for the local and nonlocal Green-function surface-integral methods. Then we address the occurrence of singularities in the integration kernels of the surface integrals. By introducing a new regularization procedure for the scattering part of the Green tensors, we extend the applicability of GSIM, both the local-response and the nonlocal-response version, to geometries where arbitrarily shaped nanowires rest on arbitrary multilayer substrates, rather than floating or hanging slightly above them.\cite{Ruppin:1992,Jung:2008}

\sssection{Surface integrals and boundary conditions}
The first numerical task of the GSIM is to solve the fields along the nanowire boundary from the following surface integrals. For the metal side of the metal-background boundary we have three surface integrals
\begin{widetext}
\begin{subequations}
\begin{eqnarray}
E_{zi}^{\rm T}(\boldsymbol \rho)&=&-\oint_{{S_{i}}} \mbox{d}\boldsymbol\rho'\,\left[e_{i}^0(\boldsymbol \rho,\boldsymbol \rho')E_{zi}^{\rm T}(\boldsymbol \rho')+e_{i}^1(\boldsymbol \rho,\boldsymbol \rho'){E_{zi,n}^{\rm T}}(\boldsymbol \rho')\right],\label{eqGIa}\\
H_{zi}^{\rm T}(\boldsymbol \rho)&=&-\oint_{{S_{i}}} \mbox{d}\boldsymbol\rho'\,\left[m_{i}^0(\boldsymbol \rho,\boldsymbol \rho')H_{zi}^{\rm T}(\boldsymbol \rho')+m_{i}^1(\boldsymbol \rho,\boldsymbol \rho'){H_{zi,n}^{\rm T}}(\boldsymbol \rho')\right],\label{eqGIb}\\
\phi_{i}(\boldsymbol \rho)&=&-\oint_{{S_{i}}}\mbox{d} \boldsymbol\rho'\,\left[p_{i}^0(\boldsymbol \rho,\boldsymbol \rho')\phi_{i}(\boldsymbol \rho')+p_{i}^1(\boldsymbol \rho,\boldsymbol \rho') \phi_{i,n}(\boldsymbol \rho')\right].\label{eqGIc}
\end{eqnarray}
\end{subequations}
The third surface integral, Eq.~(\ref{eqGIc}), is the additional one in case of nonlocal response, and is left out in the local GSIM. For the background side of the metal-background boundary we only have two surface integrals
\begin{subequations}
\begin{eqnarray}
{E_{zb}}(\boldsymbol\rho )&=&E_{zb}^{\rm inc}(\boldsymbol\rho )+\oint\limits_S \mbox{d}{\boldsymbol\rho'} \,\left[{e_{b}^0(\boldsymbol \rho,\boldsymbol \rho')}{E_{zb}}(\boldsymbol\rho ') + {e_{b}^1(\boldsymbol \rho,\boldsymbol \rho')}{E_{zb,n}}(\boldsymbol\rho ')\right]
+\oint\limits_S \mbox{d}{\boldsymbol\rho'} \,\left[{f_{b}^0(\boldsymbol \rho,\boldsymbol \rho')}{H_{zb}}(\boldsymbol\rho ') + {f_{b}^1(\boldsymbol \rho,\boldsymbol \rho')}{H_{zb,n}}(\boldsymbol\rho ')\right],\nonumber\\
\label{eqGId}\\
{H_{zb}}(\boldsymbol\rho )&=&H_{zb}^{\rm inc}(\boldsymbol\rho )+\oint\limits_S \mbox{d}{\boldsymbol\rho'} \,\left[{m_{b}^0(\boldsymbol \rho,\boldsymbol \rho')}{H_{zb}}(\boldsymbol\rho ') + {m_{b}^1(\boldsymbol \rho,\boldsymbol \rho')}{H_{zb,n}}(\boldsymbol\rho ')\right]
+\oint\limits_S \mbox{d}{\boldsymbol\rho'} \,\left[{h_{b}^0(\boldsymbol \rho,\boldsymbol \rho')}{E_{zb}}(\boldsymbol\rho ') + {h_{b}^1(\boldsymbol \rho,\boldsymbol \rho')}{E_{zb,n}}(\boldsymbol\rho ')\right].\nonumber\\\label{eqGIe}
\end{eqnarray}
\label{eqGI}
\end{subequations}
\end{widetext}
In combination with the boundary conditions
\begin{subequations}
\begin{eqnarray}
\hat n(\boldsymbol\rho)\times\mathbf E_b(\boldsymbol\rho)&=&\hat n(\boldsymbol\rho)\times\mathbf E_i(\boldsymbol\rho),\label{eqBCa}\\
\hat n(\boldsymbol\rho)\times\mathbf H_b(\boldsymbol\rho)&=&\hat n(\boldsymbol\rho)\times\mathbf H_i(\boldsymbol\rho),\label{eqBCb} \\
\epsilon_b(\boldsymbol\rho)\hat n(\boldsymbol\rho)\cdot\mathbf E_b(\boldsymbol\rho)&=&\hat n(\boldsymbol\rho)\cdot\epsilon_{\rm other}\mathbf E_{i}(\boldsymbol\rho),\label{eqBCc}
\end{eqnarray}
\label{eqBC}
\end{subequations}
for $\boldsymbol\rho$ on the boundary $S$, unique solutions of Maxwell's equations in all of space can be found. Eq.~(\ref{eqBCc}) is the additional boundary condition for nonlocal response, which is left out in the local GSIM.

In the special case of normally incident light ($k_z=0$), the above equations decouple into two independent sets. One is for TE-polarized light. In this case, the longitudinal fields can not be excited. The relevant surface integrals  are then Eqs.~(\ref{eqGIa}), (\ref{eqGId}), in combination with only the boundary conditions~(\ref{eqBCa}) and (\ref{eqBCb}). The other set is for TM-polarized light. In this case, the longitudinal fields can be excited. The required surface integrals are Eqs.~(\ref{eqGIb}), (\ref{eqGIc}), and  (\ref{eqGIe}), and all three boundary conditions in Eq.~(\ref{eqBC}) play a role.

\sssection{Singularities in integration  kernels} Some of the integration kernels in the surface integrals for nanowires have singularities, which must be treated carefully. First consider the surface integrals for the nanowires in Eqs.~(\ref{eqGIa})-(\ref{eqGIc}). There is a singularity that comes from the Green function of the Hankel-function type, which blows up in the limit $\boldsymbol\rho'\to\boldsymbol\rho$. We regularize the singularities following the routine by Garc{\' i}a de Abajo and Howie in Ref.~\onlinecite{Abajo:2002} and by Jung and S{\o}ndergaard in Ref.~\onlinecite{Jung:2008}. In particular, we note that our additional surface integral Eq.~(\ref{eqGIc}) for nonlocal response can be regularized in the same way as was known for the two others of the local GSIM,\cite{Abajo:2002,Jung:2008} because the same Green function appears in it, albeit with a different wave vector in the argument [recall Eq.~(\ref{loggeq})]. Thus the regularized version of Eq.~(\ref{eqGIc}) becomes
\begin{equation}
\frac{1}{2}\phi_{i}(\boldsymbol \rho)=-\mathcal{P}\oint_{{S_{i}}}\mbox{d} {\bm \rho'}\,\left[p_{i}^0(\boldsymbol\rho,\boldsymbol\rho')\phi_{i}(\boldsymbol \rho')+p_{i}^1(\boldsymbol\rho,\boldsymbol\rho')  \dot\phi_{i}(\boldsymbol \rho')\right],
\label{logsieqrf}
\end{equation}
where``$\mathcal{P}\oint$'' represents the integration excluding the singular point at $\boldsymbol\rho'=\boldsymbol\rho$.

\sssection{Regularization of surface integrals for background} Next we consider the surface integrals for the background in Eqs.~(\ref{eqGId}) and (\ref{eqGIe}). These integrals are the same as for local response, at least when neglecting nonlocal response in the background. Nevertheless we dwell upon them here,  because even for local response we could not find in the literature the necessary regularization procedure for touching geometries that we here present.

For the background surface integrals, both the homogenous and the scattering Green functions exhibit singularities. The  singularity associated with the homogenous Green function can be treated as above in Eq.~(\ref{logsieqrf}). Singularities associated with the scattering part of the Green function can also arise and must be treated differently. Let us first assume that the background is a slab in free space. In Eq.~(\ref{ibsicoef}), the surface integral kernels are expressed as integrals over the wavevector $k_y$, with integration limits $\pm \infty$. Singularities in the kernels may arise when the integrands in Eq.~(\ref{ibsicoef}) do not fall off rapidly enough as $k_y$ and hence $k_{\parallel}$ tend to infinity. Now in many cases singularities are prevented to occur because the reflectivities in the integrands vanish in the limit $k_{\parallel}\to \infty$. For example, $R_{\rm TE}\to 0$ as $k_{\parallel}\to\infty$, whether the slab is dielectric medium or metallic; $R_{\rm TM}\to 0$ as $k_{\parallel}\to\infty$ when the slab is composed of a  metal with nonlocal response.\cite{Ford:1984} By contrast, $R_{\rm TM}$ approaches a nonzero value as $k_{\parallel}\to\infty$ when the slab is composed of a dielectric  medium, and this case includes  a local-response metallic medium that is described by the dielectric function of the metal. This indicates that in particular TM-polarized scattering waves induced by dielectric substrates may lead to a singularity in the scattering part of the Green function.

To clearly illustrate such a scattering singularity, we consider a single nanowire resting on the $x=0$ top plane of a dielectric slab with  permittivity $\epsilon_d$ and thickness $t$. (More general substrates are discussed below.) We take the kernel $e_{0}^{bs}$ of Eq.~(\ref{eb0s}) as an example. In the limit $k_{\parallel}\to\infty$, the slab reflectivity has the value $R_{\rm TM}(\infty)=(\epsilon_d-1)/(\epsilon_d+1)$. This value is independent of the slab thickness, since waves with $k_{\parallel}\to\infty$ have an infinitely short penetration depth into the slab  and hence do not probe its thickness ($k_{x}^{2}$ approaches $-\infty$). We then split $e_{0}^{bs}$ into two parts, $e_{b}^{0s}=e_{b}^{0s1}+e_{b}^{0s2}$. In $e_{b}^{0s1}$ we deal with the possible singularity arising due to the large-$k_y$ behavior of the integrand of $e_{b}^{0s}$, whereas the integrand of $e_{b}^{0s2}$ vanishes for large $k_y$ so that $e_{b}^{0s2}$ does not have a singularity.   The possibly singular kernel term is given by
\begin{eqnarray}
e_{b}^{0s1} &=&\frac{-i k_z^2 R_{\rm TM}(\infty)}{{4\pi k_{\rho}^2  }}\int \mbox{d}{{k_y}} \frac{1}{{{k_{x}}}}\exp(i\psi)i\mathbf k_{\rho}\cdot\hat n(\boldsymbol\rho')
\nonumber\\
&=&-\frac{1}{4i}\frac{k_z^2}{k_{\rho}^2}R_{\rm TM}(\infty)\hat n(\boldsymbol\rho')\cdot\nabla_{\boldsymbol\rho'}H_0^{(1)}(k_\rho\rho_{os}),
\label{D1}
\end{eqnarray}
in terms of the angle $\psi={k_y}(y - y')-k_{x}(x+x')$, the wavevectors $k_0=\omega/c$ and $\mathbf k_{\rho}= (k_{x},k_{y})$ that satisfy the identity $k_{\rho}^2+k_z^2=k_0^2$, and the length
$\rho_{os}=\sqrt{(x+x')^2+(y-y')^2}$. The identity Eq.~(\ref{D1}) follows almost directly from the plane-wave expansion of a cylindrical wave as derived in Eq.~(2.2.11) of Ref.~\onlinecite{Chew:1990}.

A singularity of the scattering kernel $e_{b}^{0s1}$ arises when $\rho_{os}$ vanishes for a point on the surface. When does this occur? The kernel $e_{b}^{0s1}({\bm \rho},{\bm \rho'})$ appears in the surface integral~(\ref{eqGId}), and the integration runs on the surface of the nanowire. For a cylindrical nanowire this surface would be parameterized by $(x+r_{0})^{2}+y^{2}=r_{0}^{2}$.   On the outer surface of the nanowire resting on the $(x=0)$-plane, $x$ and $x'$ always have the same sign, so that $\rho_{os}$ can only vanish if $x=x'=0$. For ${\bm \rho}$ and ${\bm \rho'}$ on the circle, it follows that the scattering singularity occurs only in $(x,y)=(0,0)$, where the nanowire and the dielectric substrate touch. It holds more generally for non-cylindrical nanowires that  scattering singularities occur on the point(s) where nanowires touch  dielectric interfaces.  Following the same routine as for $e_{b}^{0s}$, the singularities in other integration kernels can also be extracted. The singularities all relate to the Hankel function, which can be treated similarly as in Eq.~(\ref{logsieqrf}). In doing so, we end up with the following
regularized background surface integrals
\begin{widetext}
\begin{subequations}
\begin{eqnarray}
S_e(\boldsymbol\rho){E_{zb}}(\boldsymbol\rho )&=&E_{zb}^{\rm inc}(\boldsymbol\rho )+\mathcal{P}\oint\limits_S \mbox{d}{\boldsymbol\rho '} \,\left[{e_{0}^b}{E_{zb}}(\boldsymbol\rho ') + {e_{1}^b}{\dot E_{zb}}(\boldsymbol\rho ')\right]+\mathcal{P}\oint\limits_S \mbox{d}{\boldsymbol\rho '} \,\left[{f_{0}^b}{H_{zb}}(\boldsymbol\rho ') + {f_{1}^b}{\dot H_{zb}}(\boldsymbol\rho ')\right],\label{E_background_regularized}\\
S_m(\boldsymbol\rho){H_{zb}}(\boldsymbol\rho )&=&H_{zb}^{\rm inc}(\boldsymbol\rho )+\mathcal{P}\oint\limits_S  \mbox{d}{\boldsymbol\rho '}  \,\left[{m_{0}^b}{H_{zb}}(\boldsymbol\rho ') + {m_{1}^b}{\dot H_{zb}}(\boldsymbol\rho ')\right]+\mathcal{P}\oint\limits_S \mbox{d}{\boldsymbol\rho '} \,\left[{h_{0}^b}{E_{zb}}(\boldsymbol\rho ') + {h_{1}^b}{\dot E_{zb}}(\boldsymbol\rho ')\right],
\label{H_background_regularized}
\end{eqnarray}
\end{subequations}
\end{widetext}
with
\begin{subequations}
\begin{eqnarray}
S_e(\boldsymbol\rho)&=&\frac{1}{2}\left[1-\frac{k_z^2}{k_{\rho}^2}R_{\rm TM}(\infty)\right], \\
S_m(\boldsymbol\rho)&=&\frac{1}{2}\left[1-\frac{k_0^2}{k_{\rho}^2}R_{\rm TM}(\infty)\right],
\end{eqnarray}\label{Se_and_Sm}
\end{subequations}
for $\boldsymbol\rho$ at the common boundary of the nanowire and the slab, and otherwise $S_e=S_m=1/2$.

Until now we have assumed that the substrate is a dielectric slab. We already discussed why the regularization procedure does not depend on the thickness of this slab. For the same reasons, we can generalize the substrate to an arbitrary multilayer dielectric. The above regularization in Eqs.~(\ref{H_background_regularized}) and (\ref{Se_and_Sm}) involves reflectivities in the limit $k_{y}\to \infty$. In this limit the reflectivity of a multilayer dielectric will be given by $R_{\rm TM}(\infty)=(\epsilon_d-1)/(\epsilon_d+1)$, where $\epsilon_d$ is to be understood as the dielectric function of the outer layer of the multilayer substrate on which the nanowires rest. Also, if the substrate is not exactly planar, then locally near the nanowire it can be approximated as planar and again the above regularization can be employed, again involving the limit reflectivity $R_{\rm TM}(\infty)$ of the dielectric material on which the nanowire rests.

By this regularization procedure we have extended the geometries to which the computationally light GSIM can be applied to experimentally relevant structures where nanowires of arbitrary shapes rest on arbitrary multilayer substrates. As will be shown in Sec.~\ref{sec:numerical_results}, it is also these touching geometries for which differences between local and nonlocal response of the plasmonic nanowires are largest.
Nanowires positioned above the substrate (i.e., non-touching geometries) are slightly simpler to analyze, because a singularity associated with the scattering part of the Green function does not arise and the above regularization is not needed.

After the above regularizations, the numerical procedure to find solutions with the local or nonlocal GSIM is as follows. By discretizing the regularized surface integrals, and using the boundary conditions, the fields along the metal-background boundaries can be solved. Then, knowing the fields on the boundaries, we can employ the surface integrals once more to obtain the fields in any position of the system, and to extract further physical quantities of our interest.

\section{Effective theory for nonlocal blueshifts}\label{sec:effective}
Before applying the hydrodynamic GSIM as developed in the previous sections, we will here give a semi-quantitative analysis of the most conspicuous optical effect of nonlocal response, namely the nonlocal blueshift of plasmonic resonance frequencies. As illustrated below, different plasmonic resonances exhibit different nonlocal blueshifts. Our analysis will explain this, and will guide our numerical investigations in Section~\ref{sec:numerical_results}.

For simplicity rather than necessity, we will neglect the dielectric response of the bound electrons in the metal, i.e., we take $\epsilon_{\rm other}=1$. This approximation is better at lower frequencies, in particular  below the band gap energy for interband transitions in the metal.

Let us consider a single subwavelength plasmonic resonator with an arbitrary shape in an inhomogeneous dielectric medium. The region of the plasmonic scatterer is denoted by $A_{\rm m}$ and has a boundary $S_{\rm m}$. First we make the usual local-response approximation, and assume that the resonator supports a SP resonance at $\omega_{\rm res}^{\rm loc}$. Neglecting loss, the equation of motion for the free electrons is $m_e({\omega_{\rm res}^{\rm loc}})^2\mathbf d=-e\mathbf E$, where $\mathbf d$ represents the displacement of the electron. The displacement $\mathbf d$ gives rise to a delta-function thin surface charge distribution $\alpha_{\rm m}$ at the boundary. This $\alpha_{\rm m}$ then induces the screening charge $\alpha_{\rm b}$ in the background.

Let us now take instead nonlocal response into account, using the same hydrodynamic model and stepwise equilibrium free-electric density for which we derived the nonlocal GSIM. Quantum spill-out and spatial variations of the equilibrium free-electric density are thus neglected, and consequently the normal component of the linear-response free-electron current vanishes on the metal-dielectric boundary. We then find that  the hydrodynamic pressure gradient smears out the linear-response surface charge $\alpha_{\rm m}$ into a surface charge distribution of finite thickness, decaying away from the surface and {\em into} the metal approximately exponentially as $\exp(-k^{\rm L}\Delta)$, where $k^{\rm L}$ is the longitudinal wavenumber, and $\Delta$ the distance to the boundary. Thus, rather than exactly on the surface $S_{\rm m}$ as for local response, the nonlocal surface charge is effectively accumulated on a smaller boundary $S_{\rm m}'$, at a distance of $1/k^{\rm L}$ within $S_{\rm m}$. In the region inside $S_{\rm m}'$, denoted by $A_{\rm m}'$, the electric field $\mathbf E'$ is enhanced owing to the inward displacement of the surface charge by the pressure-gradient force. For $A_{\rm m}'$ to exist, we must of course require that the surface layer thickness $1/k^{\rm L}$ is smaller than the effective radius. We furthermore assume that the free-electron displacement $\mathbf{d}$ within $A_{\rm m}'$ is unchanged.
Then, we approximately have $m_e({\omega_{\rm res}^{\rm nloc}})^2\mathbf d=-e\mathbf E'$, where $\omega_{\rm res}^{\rm nloc}$ represents the new resonance frequency modified by the nonlocal response. This explains that the nonlocal response indeed blueshifts the resonance frequency, i.e., $\omega_{\rm res}^{\rm nloc}>\omega_{\rm res}^{\rm loc}$. Moreover, we can also understand the blueshift quantitatively. By relating $\omega_{\rm res}^{\rm nloc}$ to $\omega_{\rm res}^{\rm loc}$ by integrating the free-electron equation of motion in the area $A_{\rm m}'$, we find the approximate relation
\begin{equation}
\omega_{\rm res}^{\rm nloc}\approx\omega_{\rm res}^{\rm loc}\left(\frac{{\int_{{A_m'}} \mbox{d}{\mathbf r|\mathbf E'|^2} }}{{\int_{{A_m'}} \mbox{d}{\mathbf r|\mathbf E|^2} }}\right)^{1/4},
\label{blueshift1}
\end{equation}
which we will test in Sec.~\ref{sec:numerical_results} using our nonlocal GSIM.

As a specific example that allows analytical treatment, let us now consider a subwavelength metallic cylinder with radius $r_0$ in a homogenous dielectric background, and use Eq.~(\ref{blueshift1}) to derive the resonance frequency $\omega_{\rm res}^{\rm nloc}$.

In local response and in the quasi-static limit, it is well known that the cylinder supports a SP resonance at the frequency $\omega_{\rm res}^{\rm loc}$ for which $\epsilon_{\rm m}^{\rm T}(\omega_{\rm res}^{\rm loc})$ equals $-\epsilon_{\rm b}$. At the boundary, the surface charge $\alpha=\alpha_{\rm m}+\alpha_{\rm b}$ is accumulated such that $\alpha_{\rm b}/\alpha_{\rm m}=(\epsilon_{\rm m}^{\rm T}\epsilon_{\rm b}-\epsilon_{\rm m})/(\epsilon_{\rm b}-\epsilon_{\rm m}^{\rm T}\epsilon_{\rm b})$. The electric field $\mathbf E$ in the metal is the sum of two terms, $\mathbf{E}=\mathbf{E}_{\rm m}+\mathbf E_{\rm b}$, where $\mathbf{E}_{\rm m,b}$ are due to the charge densities $\alpha_{\rm m,b}$, respectively, which control their relative magnitude by $|\mathbf{E}_{\rm m}|/|\mathbf E_{\rm b}|=\alpha_{\rm m}/\alpha_{\rm b}$.

Turning from local to nonlocal response, the charge density $\alpha_{\rm m}$ is effectively distributed on a smaller surface, as discussed above, characterized by a smaller radius $r_0'$ equal to $(r_0-1/k^{\rm L})$. We thus have $k^{\rm L} r_0 >1$ as the consistency requirement for our effective theory. The smaller effective radius results in an enhancement of $\mathbf{E}_{\rm m}$ approximately by a factor of $(r_0/r_0')^l$, where $l$ represents the angular momentum of the SP mode. For example, $l=1$ for the dipole mode, $l=2$ for the quadrupole mode, etc. Furthermore, the surface charge density $\alpha_{\rm b}$ is reduced by a factor of $(r_0'/r_0)^l$, and $\mathbf E_{\rm b}$ is correspondingly reduced by the same factor. Taking these effects together, and assuming that nonlocal effects are small, the total electric field in the metal is approximately enhanced  by a factor $1+\epsilon_{\rm b}l/(k^{\rm L} r_0)$. Inserting these results into Eq.~(\ref{blueshift1}), the nonlocal resonance frequency $\omega_{\rm res}^{\rm loc}$ is found to be
\begin{equation}
\omega_{\rm res}^{\rm nloc}\approx\omega_{\rm res}^{\rm loc}\left( 1+\frac{\epsilon_bl}{2 k^{\rm L} r_0}\right).\hspace{7mm}{\rm (cylinder)}
\label{resf1}
\end{equation}
Thus the magnitude of the nonlocal blueshift essentially depends on four parameters, namely the longitudinal wavevector $k^{\rm L}$,
the particle size $r_0$, the angular-momentum number $l$ of the resonant SP mode under consideration, and finally the background dielectric function $\epsilon_{\rm b}$. A higher $\epsilon_{\rm b}$ gives rise to a larger blueshift.\cite{Raza:2012b} Of the two parameters of the nanowire $k^{\rm L}$ and $r_0$, the former is determined by the intrinsical nonlocal $\beta$ factor and the operating frequency, while the latter can be experimentally varied. Smaller plasmonic resonators give rise to a larger nonlocal blueshift, as is well known. Less well known, although seen but not analyzed in early calculations for a sphere in a homogeneous background,\cite{Boardman:1977} is our important point that the nonlocal blueshift grows with the angular momentum of the SP mode. In general, mode profiles corresponding to higher values of $l$ show a stronger spatial variation.

By the same approach as leading to Eq.~(\ref{resf1}), the nonlocal resonance frequency for a plasmonic sphere can be derived as
\begin{equation}
\omega_{\rm res}^{\rm nloc}\approx\omega_{\rm res}^{\rm loc}\left[1+\frac{\epsilon_b(l+1)}{2 k^{\rm L} r_0}\right].\hspace{7mm}{\rm (sphere)}
\label{resf2}
\end{equation}
From Eqs.~(\ref{resf1}) and~(\ref{resf2}) we find that the relative blueshift $(\omega_{\rm res}^{\rm nloc}-\omega_{\rm res}^{\rm loc})/\omega_{\rm res}^{\rm loc}$ depends on the nature of the plasmonic resonance and grows linearly with the angular momentum number $l$, both for cylinders and for spheres. We will illustrate this angular momentum dependence by numerically exact calculations below. Furthermore, this dependence could be tested experimentally.
Clearly, from the factors $(l+1)$ in Eq.~(\ref{resf2}) and $l$ in Eq.~(\ref{resf1})  it follows that a 3D plasmonic sphere is more sensitive to the nonlocal response than a 2D wire. The relative difference in their blueshifts is a factor of two for the dipole resonance with $l=1$, and approaches unity  for high-order resonances.

Besides the $l$-dependence, our second important point is that Eq.~(\ref{resf1}) leads to some useful insights about blueshifts of plasmonic nanowires in  {\em inhomogeneous} backgrounds, even though it was derived for a homogeneous background. For example, if we embed a cylindrical plasmonic nanowire on a substrate, then as a result it will typically exhibit a more inhomogeneous and more confined mode profile.\cite{Jung:2011,Zhang:2012} Consequently the expansion of the surface charge density into angular-momentum eigenmodes will show a larger contribution from larger angular momenta. Based on the effective theory developed here and in particular in Eq.~(\ref{resf1}), we expect an accordingly larger nonlocal blueshift. We will quantify and verify this idea in Section~VII, using our numerically exact nonlocal GSIM method developed in Sec.~\ref{sec:GSIM}.


\section{Numerical analysis: nanowire with and without a substrate}\label{sec:numerical_results}

In this section we employ the hydrodynamic  GSIM as developed in Sec.~\ref{sec:GSIM} to numerically investigate the optical effects of nonlocal response in plasmonic nanowires. The important physical quantity considered here is the extinction cross section $\sigma _{\rm ext}$, which in general is defined as the sum of the absorption and scattering cross sections,
\begin{equation}\label{sigmaextdef}
\sigma _{\rm ext}=\sigma_{\rm abs}+\sigma_{\rm sca}.
\end{equation}
Cross sections are usually defined as an area, but for the infinitely long nanowires that we consider here, we will instead consider cross sections per length unit of the nanowire, with the dimension of a length. So let us now introduce the cross sections $\sigma_{\rm abs}$ and $\sigma_{\rm sca}$.

We consider TM-polarized incident plane waves with $H_{zb}^{\rm inc}=\exp(ik_0x)$ and $E_{zb}^{\rm inc}=0$.
The ratio between the  electromagnetic power that is absorbed by the nanowire and the incident electromagnetic power of the plane wave is known as the absorption cross section $\sigma_{\rm abs}$, which can be expressed as the surface integral
\begin{equation}
\sigma _{\rm abs} = \oint_{S}\mbox{d}\boldsymbol\rho\,{\operatorname{Re}\left(\frac{i H_{zb}{H}_{zb,n}^*}{k_b|H_{zb}^{\rm inc}|^2}  \right)},
\end{equation}
where the superscript ``$*$'' represents the complex conjugate operation. Likewise, the scattering cross section $\sigma_{\rm sca}$ is defined as the ratio between the  electromagnetic power that is scattered by the nanowire and the incident power. In a lossless homogenous background, such as free space, the scattered power can be expressed in terms of only the far-field radiation power. By contrast, in an inhomogeneous and lossy background, the scattered power includes besides the far-field radiation power both the localized waveguide power and the power absorbed by the lossy background. The scattering cross section $\sigma_{\rm sca}$ can also be written as a surface integral,
\begin{equation}\label{sigmascadef}
\sigma _{\rm sca} = -\oint_{S}\mbox{d}\boldsymbol\rho\,{\operatorname{Re}\left(\frac{i H_{zb}^s{ H}_{zb,n}^{s*}}{k_b|H_{zb}^{\rm inc}|^2}  \right)},
\end{equation}
where $H_{zb}^s$ represents the scattered field defined as $H_{zb}^s=H_{zb}-H_{zb}^{\rm inc}$, and ${H}_{zb,n}^s= H_{zb,n}-{H}_{zb,n}^{\rm inc}$.

Below we present calculations of extinction cross sections, first for free-standing plasmonic nanowires in Sec.~\ref{SEC:FREESTANDING}, then for nanowires above a substrate in Sec.~\ref{SEC:wire_above_substrate}, and finally in Sec.~\ref{sec:wire_on_substrate} for nanowires resting directly on a substrate.

\subsection{Free-standing nanowire}\label{SEC:FREESTANDING}
The extinction cross section of a free-standing nanowire can be computed analytically, not only in local response but also in the hydrodynamic Drude model.\cite{Ruppin:2001,Raza:2011} It is thus an excellent benchmark problem for numerical methods. In fact, this same benchmark problem was used independently by two groups to show the accuracy of their finite-element method implementations of the hydrodynamic model.\cite{Toscano:2012a,Hiremath:2012} Here we put our nonlocal GSIM to the same test.

We consider an Au cylinder with a radius $r_0$ in a free-space background. An exact solution can be found by a nonlocal extension of Mie scattering theory.\cite{Ruppin:2001,Raza:2011,Raza:2012a} We use the following parameters for gold: $\hbar \omega _{\rm {pf}}= 8.812 \rm{eV}$, $\hbar {\gamma_{\rm f}} = 0.0752 \rm{eV}$, and $v_{\rm F}=1.39\times10^6\,\rm{m/s}$. As in  Sec.~\ref{sec:effective}, for simplicity we neglect the contribution of the bound electrons, i.e. we take $\epsilon_{\rm other}=1$.

\begin{figure}[t]
\includegraphics[width=0.45\textwidth]{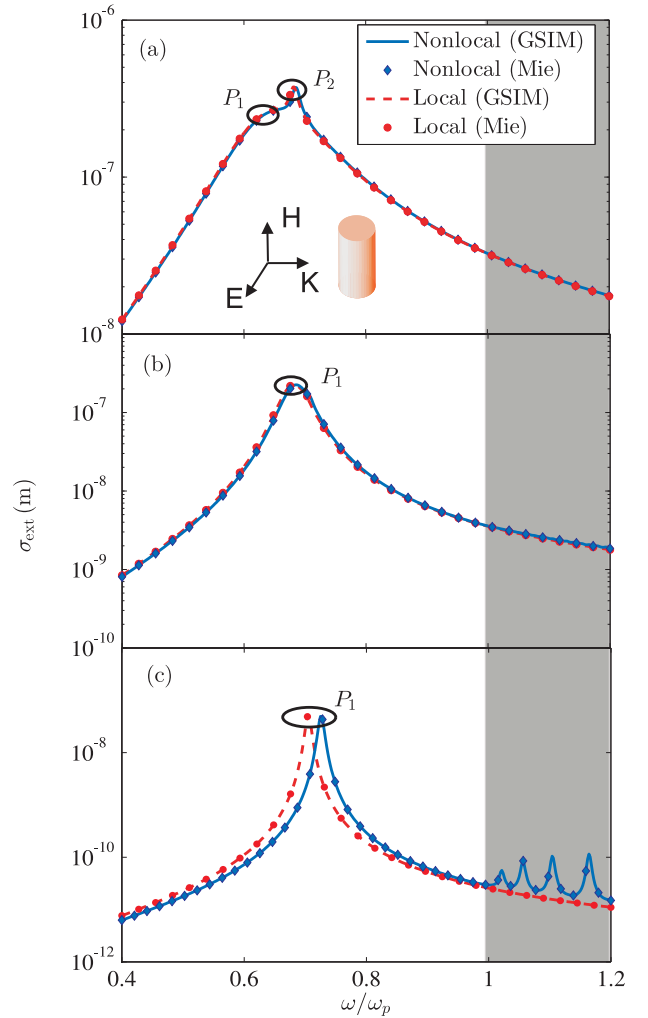}
\caption
{Extinction cross section of an Au cylinder in a free-space background, for a TM-polarized incident plane wave. The cylinder is described both by the usual local-response model and by the hydrodynamic Drude model. The simple wire geometry serves as an excellent benchmark problem: analytically exact calculations (local and nonlocal Mie theory) are compared with a numerically exact method (local and our nonlocal GSIM). The cylinder radius is (a) 20$\rm nm$; (b) 10$\rm nm$; and (c) 2$\rm nm$.}\label{Fig:freestanding}
\end{figure}
In Figs.~\ref{Fig:freestanding}(a)-(c), the extinction cross section curves are depicted for the local model as well as the nonlocal HDM, by using both the GSIM and Mie's scattering theory. Clearly, the results from two different methods agree very well with each other. This verifies the validity of the GSIM as a numerically exact method, both for local response (as is known in the literature) and for nonlocal response (which is our new result). For nonlocal response the benchmark is more stringent, since not only the blueshifted resonances should come out right with the nonlocal GSIM, but also the series of hydrodynamic resonances above the plasma frequency. And indeed they do.

In Fig. 2(a), two peaks $P_1$ and $P_2$ are observed below the plasma frequency. They correspond to the dipole and quadrupole resonances, respectively. $P_1$ is broader than $P_2$ because the dipole resonance is more radiative than the quadrupolar one. Comparing the local and nonlocal curves, their resonance frequencies are nearly the same, because nonlocal response has a weak effect on the structure that is much larger than the Thomas--Fermi screening length.\cite{Ruppin:2001,Raza:2011,Toscano:2012a} For Fig.~\ref{Fig:freestanding}(b) we reduce $r_0$ to $10\,\rm nm$ and observe that the $P_2$ disappears from the extinction cross sections, exemplifying that the quadrupole resonance becomes harder to excite by a plane wave as the nanowire becomes smaller. Furthermore, $P_1$ is narrower than in Fig.~\ref{Fig:freestanding}(a) because the smaller scatterer is less radiative. A tiny nonlocal blueshift of $P_1$ becomes just visible, and it is indeed known that nonlocal blueshifts increase as the size $r_0$ decreases. Here for $r_0=10\,\rm nm$, the relative blueshift $\Delta\omega_{\rm res}\equiv (\omega_{\rm res}^{\rm nloc}-\omega_{\rm res}^{\rm loc})/\omega_{\rm res}^{\rm loc}$ is a mere $0.6\%$. In Fig.~\ref{Fig:freestanding}(c), we take $r_0=2\,\rm nm$. The nonlocal response blueshift for $P_1$ resonance becomes clearly noticeable with $\Delta\omega_{\rm res}\approx 3\%$. Additionally, a series of peaks corresponding to the optical excitations of resonant longitudinal modes appear above the plasma frequency,\cite{Ruppin:2001,Raza:2011,Toscano:2012a} the so-called unusual resonances.\cite{Raza:2011} By contrast,  these longitudinal modes were not visible in Figs. 2(a) and (b), because the frequency spacing between the different longitudinal modes becomes smaller for larger nanowires, and the damping loss smears out these modes.

\begin{figure}[t]
\includegraphics[width=0.45\textwidth]{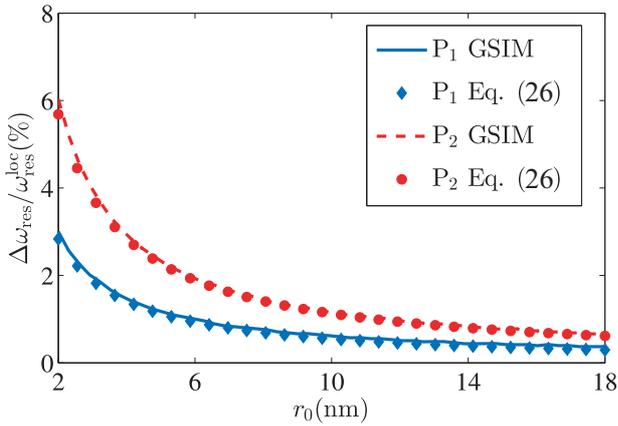}
\caption
{The nonlocal blueshift $\Delta\omega_{\rm res}$ of the dipole and quadrupole resonances for an Au cylinder in free space, as a function of cylinder radius $r_{0}$.}
\label{Fig:blueshift_benchmark}
\end{figure}

To test the effective theory  of Sec.~\ref{sec:effective}, in Fig.~\ref{Fig:blueshift_benchmark}
%
we show the relative blueshift $\Delta\omega_{\rm res}$ as a function of $r_0$,  both for the dipole and the quadrupole resonances of the free-standing nanowire. To excite the quadrupole resonance $P_2$, we used a cylindrical wave with angular momentum $l=2$ as the incident wave. The relative blueshifts $\Delta\omega_{\rm res}$ are calculated twice, using our numerically exact GSIM and our approximate expression Eq.~(\ref{resf1}). Clearly, the results from two methods are in good agreement, which is a first test of the validity of the effective theory in Section~\ref{sec:effective}. The information contained in Fig.~\ref{Fig:blueshift_benchmark} is twofold: (i) the nonlocal blueshift increases for smaller radius $r_0$, as was known before~\cite{Raza:2011}; (ii) the relative nonlocal blueshift for the quadrupole resonance is indeed two times larger than that for the dipole resonance, in agreement with Eq.~(\ref{resf1}). This significant $l$-dependence a new result. Our numerically exact GSIM confirms that higher-order SP resonances are significantly more sensitive to nonlocal response.

\subsection{Nanowire above a dielectric substrate}\label{SEC:wire_above_substrate}
Let us now consider the extinction cross section of a plasmonic nanowire positioned at a finite height above a semi-infinite dielectric substrate, and investigate the interactions between wire and substrate. The cross sections are computed using Eqs.~(\ref{sigmaextdef}-\ref{sigmascadef}). In comparison to other methods such as the finite-element method, the unique advantage of the GSIM is that even infinitely long and thick substrates can be taken exactly into account, by using the exact background Green function (given in Appendix~\ref{Sec:slab_background}). This advantage of the GSIM was stressed in Ref.~\onlinecite{Jung:2008} for local response, and here we illustrate that the nonlocal GSIM has the same advantage.

\begin{figure}[t]
\includegraphics[width=0.45\textwidth]{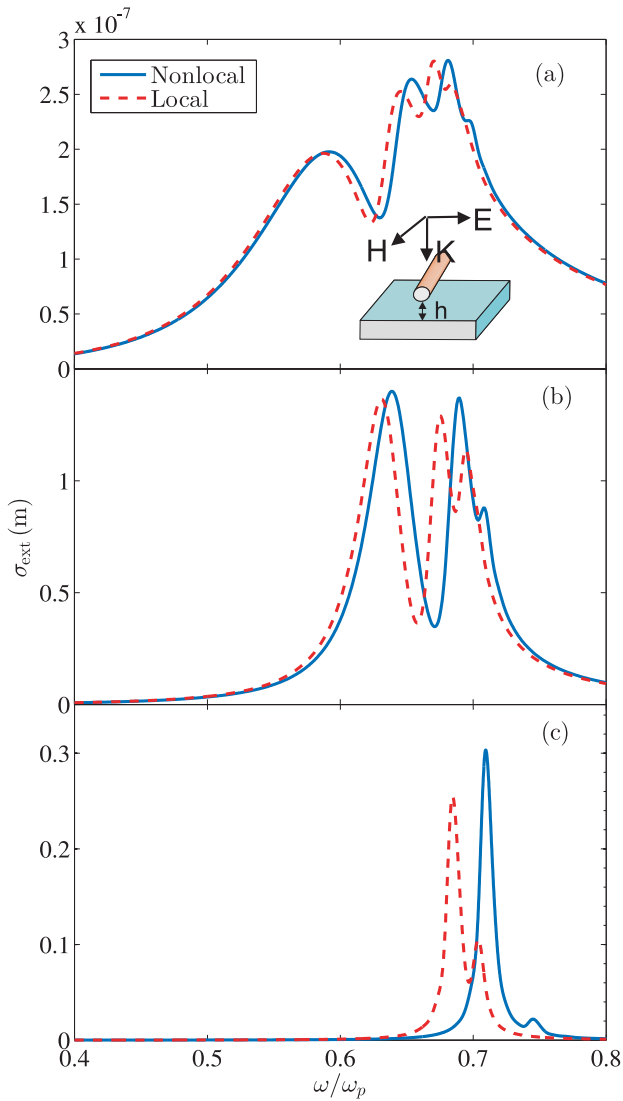}
\caption
{The extinction cross section of an Au cylindrical nanowire positioned above a dielectric substrate of refractive index $1.5$, for a TM-polarized plane wave incident from the top. The distance between the Au cylinder and the substrate is $h=1\,{\rm nm}$. The wire radius $r_{0}$ is (a) 20$\,\rm nm$; (b) 10$\,\rm nm$; and (c) 2$\,\rm nm$. Local and nonlocal results are calculated with standard local GSIM and with our generalized nonlocal GSIM, respectively.}\label{fig:wireandsubstrateh1}
\end{figure}

\sssection{Substrate enhances nonlocal blueshift} In Fig.~\ref{fig:wireandsubstrateh1}(a)-(c), we present extinction cross sections of the same three nanowires as in Fig.~\ref{Fig:freestanding}, but now positioned a single nanometer above a dielectric substrate with refractive index $1.5$.
In comparison to Fig.~\ref{Fig:freestanding}, both local and nonlocal resonances are now redshifted. A simple explanation is that the substrate increases the average background permittivity of the wire, and  plasmonic nanowires in homogeneous backgrounds with higher permittivities have lower resonance frequencies.\cite{Raza:2012b}
Alternatively, the redshift could also be explained from the hybridization theory by considering the interactions between the nanowire and its electromagnetic image induced by the substrate.\cite{Toscano:2012a,Kottmann:2001,Nordlander:2003,Nordlander:2005,Davis:2010}
For the hydrodynamic Drude model, in Fig.~\ref{fig:wireandsubstrateh1} again blueshifts are observed with respect to the local-response resonances. In particular, for the first-order SP resonance mode, we find the relative blueshifts $\Delta\omega_{\rm res}\approx 0.7\%$ for $r_0=20\,\rm {nm}$, and $\Delta\omega_{\rm res}\approx 1.25\%$ for $r_0=10\,\rm {nm}$, and finally $\Delta\omega_{\rm res}\approx 3.6\%$ for $r_0=2\,\rm {nm}$. These relative blueshifts are larger than those of Sec.~\ref{SEC:FREESTANDING} without the substrate, so bringing a substrate close to nanowires enhances their nonlocal blueshifts. This is interesting and of practical importance for the interpretation of experiments. For example for the EELS experiments on few-nanometer sized plasmonic spheres on substrates of Ref.~\onlinecite{Raza:2012b}, where larger blueshifts were observed than calculated hydrodynamic blueshifts for free-standing nanospheres.

\begin{figure}[t]
\includegraphics[width=0.45\textwidth]{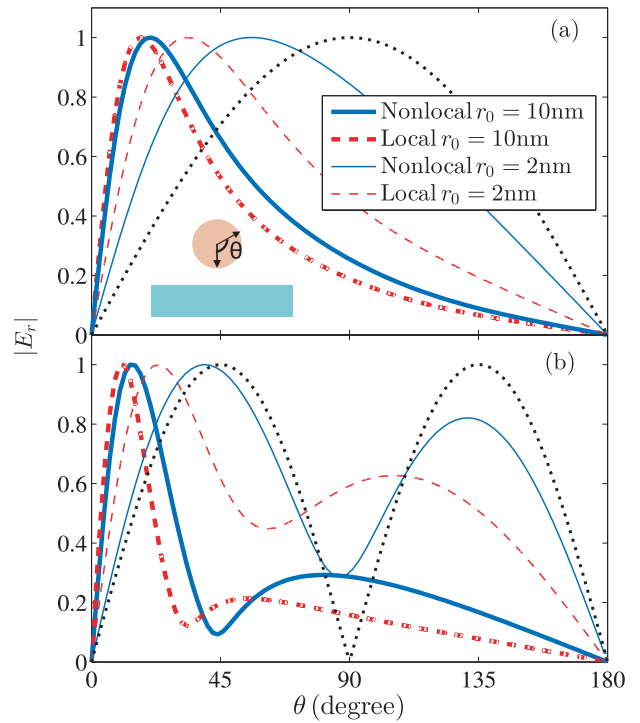}
\caption
{For the wire with $h=1\,{\rm nm}$ above the substrate as shown in the inset of Fig.~\ref{fig:wireandsubstrateh1}, the radial component of the electric field (scaled with respect to its maximal value) along the nanowire boundary of the background side for (a) the first-order SP resonance mode and (b) the second-order SP resonance mode, calculated with the local GSIM as well as the nonlocal GSIM. The dotted curves in (a) and (b) represent the pure dipole and quadrupole modes, respectively, for the same nanowire but without a substrate.}\label{Fig:Enormaltheta}
\end{figure}

\sssection{Qualitative explanation of larger blueshift}
To qualitatively explain why the substrate enhances the nonlocal effects, we boldly apply Eq.~(\ref{resf1}) that was orginally derived for homogeneous backgrounds to inhomogeneous ones. According to Eq.~(\ref{resf1}), the background may affect the nonlocal blueshift through effectively modifying the background dielectric function $\epsilon_b$ and the angular momentum $l$ associated with the resonance. First, in the presence of the substrate one can interpret $\epsilon_b$ as an average value, which characterizes the screening charge contribution from the substrate. Independent of how this average is computed, this average value goes up when adding the substrate to the initial free-space environment. Eq.~(\ref{resf1}) then tells that the nonlocal blueshift of the plasmonic wire's resonance increases. Second, regarding the angular-momentum parameter $l$, the substrate breaks the symmetry of the background, and makes the mode profile more inhomogeneous, as is known for local response. To illustrate this both for local and nonlocal response, we plot the radial component of the outer electric field along the nanowire boundary for the first- and second-order resonance modes in Fig.~\ref{Fig:Enormaltheta}(a) and (b), respectively.
Despite the nearby substrate, these modes are still similar to the pure dipole and quadrupole modes in the absence of a substrate. The radial components $E_{\rm r}$ for local and nonlocal response differ considerably for all angles, and one reason is the additional boundary condition~(\ref{ABCintermsofE}) for nonlocal response. However, the important point is that all field distributions shown in Fig.~\ref{Fig:Enormaltheta} are more concentrated on the substrate side, and in that sense are more inhomogeneous. This implies that the effective angular momenta of the lowest two resonances modes should be larger than $l=1$ and $2$, respectively. Based on the hydrodynamic Drude model and in particular on Eq.~(\ref{resf1}), one therefore expects concomitant larger nonlocal blueshifts due to the presence of the substrate. Thus we can indeed qualitatively explain that the substrate enhances the nonlocal blueshift. It does so by increasing both the average $\epsilon_b$ and the effective $l$.

\sssection{Quantitative explanation of larger blueshift}
To support the above qualitative arguments by numbers, we will quantify how the substrate modifies the parameters $\epsilon_b$ and $l$. We first define the {\em effective angular momentum} $l_{\rm eff}$ for an arbitrary mode profile in the local-response approximation, by expanding its associated surface charge on the metal surface $\alpha_{\rm m}({\boldsymbol\rho})$ in cylindrical harmonics, with weights $\alpha_{\rm ml}$. The derivation can be found in Appendix~\ref{Sec:l_eff_def}, and for local response the result is
\begin{equation}
{l_{\rm eff}}^{-1} = \frac{{\sum\limits_{l\ne 0}}|{\alpha_{\rm ml}}{|^2}{l}^{-1} }{{\sum\limits_{l\ne 0}} {|{{\alpha_{\rm ml}}}{|^2}} }.
\label{eqeffl}
\end{equation}

\begin{figure*}[t]
\includegraphics[width=0.8\textwidth]{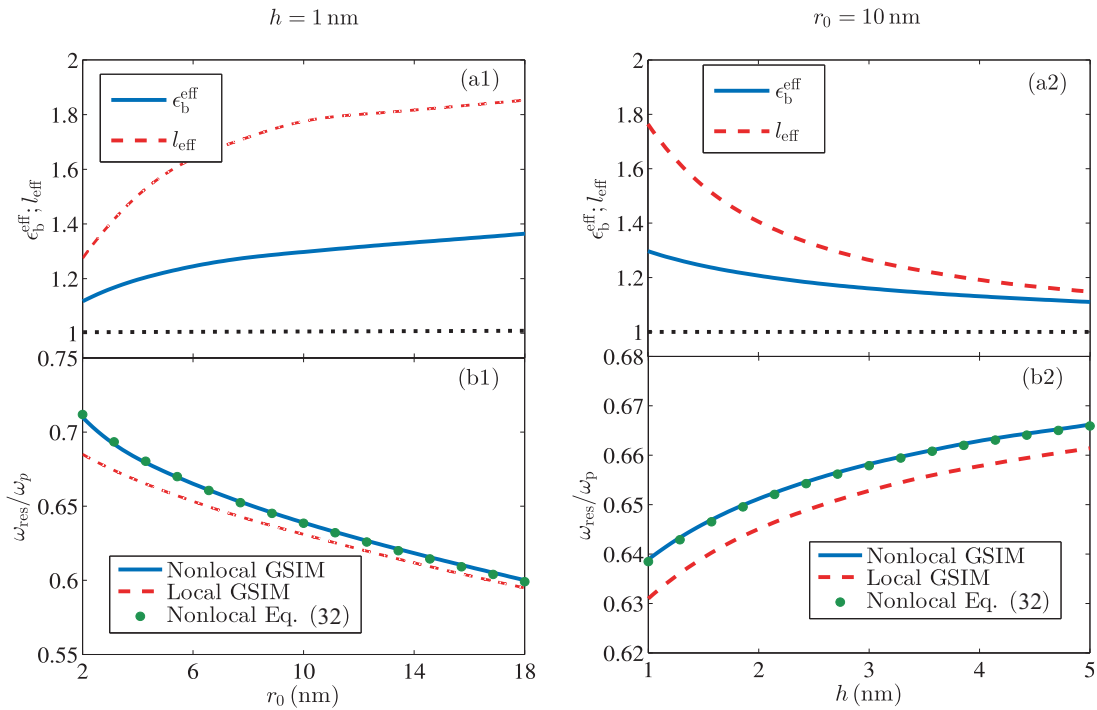}
\caption{For the wire above the substrate as shown in the inset of Fig.~\ref{fig:wireandsubstrateh1}, in panels (a1) and (b1) we keep the wire-substrate distance fixed at $h=1\,{\rm nm}$ and vary the wire radius $r_{0}$, while in panels (a2) and (b2) we keep the radius fixed at $10\,{\rm nm}$ and vary its distance to the substrate. For the first-order resonance, panels (a1,a2) show effective angular momenta and average background dielectric functions in nonlocal response, while panels (b1,b2) show scaled resonance frequencies both in local and nonlocal response. The dot curves in (a1) and (a2) correspond to the first-order (pure dipole) resonance in nanowires without the substrate.
}\label{fig:leff_average_epsilon}
\end{figure*}

In the case without the substrate, Eq.~(\ref{eqeffl}) reproduces the exact angular momenta $l=1$ and $l=2$ for the dipole and quadrupole modes, respectively. With the substrate as in Fig.~\ref{fig:wireandsubstrateh1}, we numerically calculate with Eq.~(\ref{eqeffl}) the effective angular momentum for the first-order resonance mode, as shown in Fig.~\ref{fig:leff_average_epsilon}(a1).
As expected, we see that $l_{\rm eff}>1$ and increases with $r_0$. This is consistent with the field distributions in Fig.~\ref{Fig:Enormaltheta}(a), where the field is more enhanced on the substrate side for $r_0=10\,{\rm nm}$ than for $2\,{\rm nm}$.

Next, let us consider how one could define the {\em effective background permittivity} $\epsilon_b^{\rm eff}$. As seen in Fig.~\ref{Fig:Enormaltheta}, the field distributions of the first- and second-order modes still resemble the pure dipole and quadrupole resonance modes of free-standing nanowires. We define the effective permittivity $\epsilon_b^{\rm eff}$ as the homogeneous background permittivity around the nanowire that would produce the same local resonance frequency as does the nanowire in the inhomogeneous background. The same definition  was used in  Ref.~\onlinecite{Scholl:2012}.

In Fig.~\ref{fig:leff_average_epsilon}(a1), where the nanowire system is as in Fig.~\ref{fig:wireandsubstrateh1}, we plot $\epsilon_b^{\rm eff}$ for the first-order resonance mode. Clearly, $\epsilon_b^{\rm eff}$ is larger than unity, the value in the absence of the substrate. It is important to notice that $\epsilon_b^{\rm eff}$ increases approximately by  $20\%$ as  the nanowire radius $r_0$ grows from $2\,{\rm nm}$ to $18\,{\rm nm}$. Despite the different geometries and materials considered, this increase is somewhat in conflict with  Ref.~\onlinecite{Scholl:2012}, where in the analysis of EELS experiments on spheres on supporting thin substrates, it was assumed that $\epsilon_b^{\rm eff}$ is independent of the sphere radius. Panel~\ref{fig:leff_average_epsilon}(a2) shows a weaker dependence of $\epsilon_b^{\rm eff}$ on the wire-substrate distance $h$, at least for $h>1\,{\rm nm}$. The case of wires touching the substrate ($h=0\,{\rm nm}$) will be addressed in Sec.~\ref{sec:wire_on_substrate}.

We defined $l_{\rm eff}$ to quantify effects of the inhomogeneity of the substrate, whereas we defined $\epsilon_{\rm eff}$ assuming that the background can be described as an effectively homogeneous one. There is no real contradiction here and the results that we obtain are accurate as long as nonlocal blueshifts are small perturbations, as we shall see.
To prove that the definitions of $\epsilon_b^{\rm eff}$ and $l_{\rm eff}$ make good physical sense, also in combination, we define the {\em effective relative nonlocal blueshift} $\omega_{\rm res}^{\rm nloc, eff}$ based on the expression Eq.~(\ref{resf1}) for nonlocal blueshift in a homogeneous medium, as
\begin{equation}\label{effective_relative_nonlocal_blueshift}
\Delta\omega_{\rm res}^{\rm nloc, eff} = \frac{\epsilon_b^{\rm eff} l_{\rm eff}}{2 k^{\rm L} r_{0}},
\end{equation}
thus simply replacing $l$ and $\epsilon_{b}$ in Eq.~(\ref{resf1}) by their effective values as defined above. We calculate this dimensionless blueshift for the first-order SP mode in Fig.~\ref{fig:leff_average_epsilon}(b1) and (b2), varying $r_{0}$ and $h$, respectively. In the same panels the numerically exact resonances $\omega_{\rm res}^{\rm nloc}$ and $\omega_{\rm res}^{\rm nloc}$ are also shown. It is seen in Figs.~\ref{fig:leff_average_epsilon}(b1) and (b2) as one of our main results that the effective nonlocal blueshift~(\ref{effective_relative_nonlocal_blueshift}) agrees quite well with the numerically exact value in the large and physically relevant parameter range where the nonlocal blueshift stays within a few percent. This confirms that our definitions of the effective parameters $\epsilon_b^{\rm eff}$ and $l_{\rm eff}$ are useful, and that we can apply our effective theory to explain nonlocal blueshifts of nanoplasmonic wires even in inhomogeneous backgrounds.

\subsection{Nanowire on a dielectric substrate}\label{sec:wire_on_substrate}
\begin{figure}[t]
\includegraphics[width=0.45\textwidth]{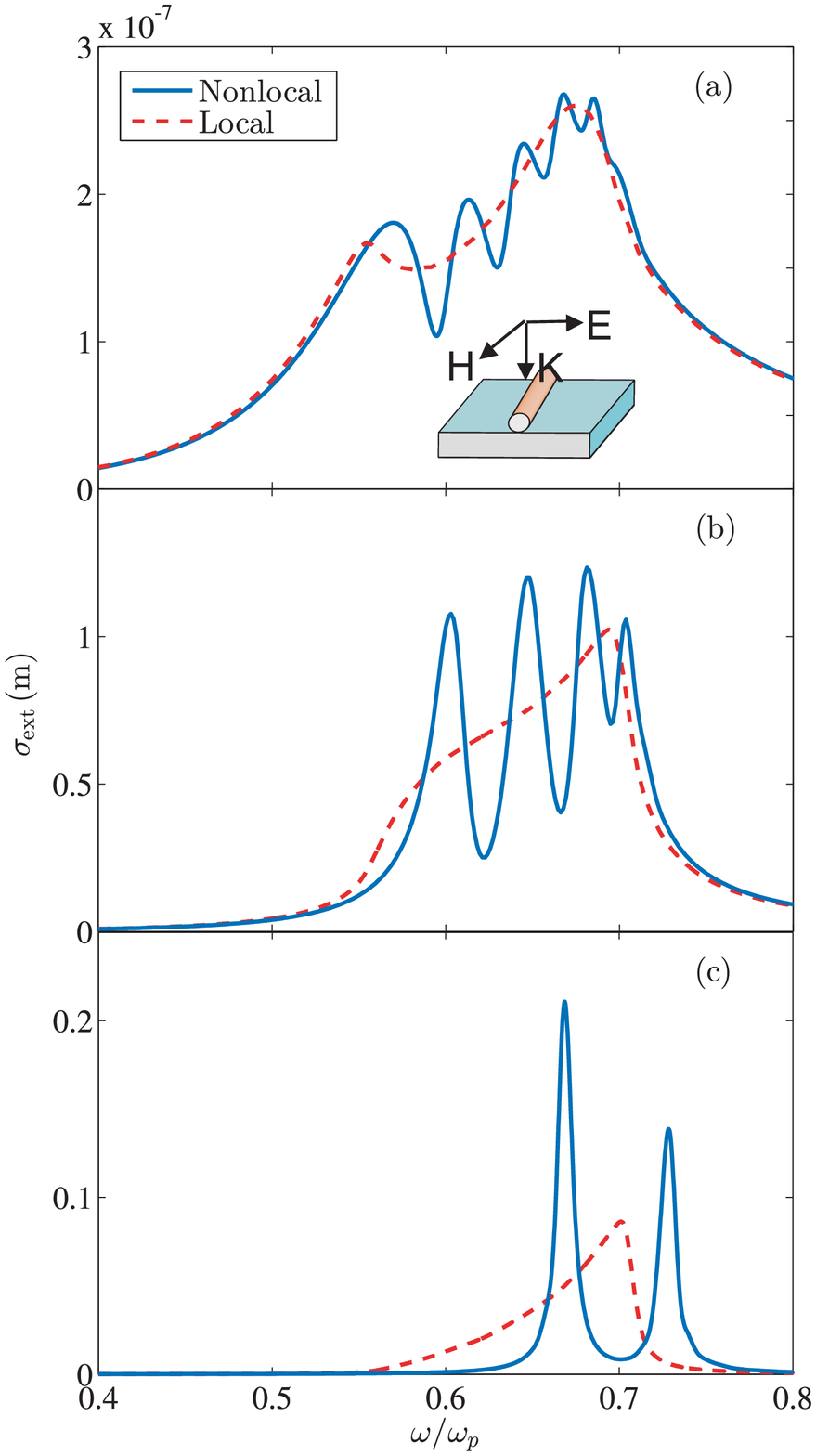}
\caption
{The extinction section for an incident TM-polarized plane wave of a gold cylindrical nanowire resting directly ($h=0$) on a dielectric substrate of refractive index $1.5$. The nanowire radii are (a) 20$\rm nm$; (b) 10$\rm nm$; and (c) 2$\rm nm$.}\label{fig:wire_touches_substrate_resonances}
\end{figure}

Having studied nanowires without substrate and above a substrate, in this subsection we consider nanowires directly  on a dielectric substrate,  i.e. the touching geometry with $h=0$. This is the first geometry in this paper for which the new regularization of the scattering Green functions in the GSIM of Sec.~\ref{Sec:numerical} is required. It is required both for the local and for the nonlocal GSIM. In Fig.~\ref{fig:wire_touches_substrate_resonances}(a)-(c),
extinction curves are depicted for the same three wire radii as before. The substrate is also the same as in Fig.~\ref{fig:wireandsubstrateh1}. However, this time the local extinction curves and their nonlocal counterparts show completely different features. This is quite unlike the free-standing nanowire  in Fig.~(\ref{Fig:freestanding}) and the wire above the substrate in Fig.~(\ref{fig:wireandsubstrateh1}), where the nonlocal response as a small perturbation only modifies the local curves slightly. The brief explanation is that the SP mode in the local description diverges in the limit of vanishing gap size between the nanowire and the dielectric substrate, whereas in the nonlocal HDM, no such divergence occurs.

Analogous large differences between local and nonlocal response for touching plasmonic nanoparticles have been predicted for the absorption cross section of two touching plasmonic spheres already by Fuchs and Claro in Ref.~\onlinecite{Fuchs:1987}. Recently Fern{\'a}ndez-Dom{\'{\i}}nguez {\em et al.} elegantly combined transformation optics with the hydrodynamical model to calculate the field enhancement near two touching plasmonic nanowires.~\cite{FernandezDominguez:2012} The general picture is that upon reducing the distance from 1 nm down to zero, the local-response resonances vary wildly even in the final {\AA}ngstrom distance, whereas the nonlocal-response resonances ``freeze out''. Our Fig.~\ref{fig:wire_touches_substrate_resonances} illustrates that one does not need plasmonic dimers to see such large differences between local and nonlocal response in the (almost) touching geometry, since a single plasmonic nanowire above/on a planar dielectric substrate suffices.

To better understand the differences for local and nonlocal response for our geometry, we model the surface plasmons in the (near-)touching region as those of a planar metal-air-dielectric structure.  In the quasi-static limit the dispersion relation of the local SP mode supported by the metal-air-dielectric sandwich structure is $\tanh ({k_{\rm sp}} h) =  - ({\epsilon _{\rm d}} + \epsilon _{\rm m}^{\rm T})/(1 + {\epsilon _{\rm d}}\epsilon _{\rm m}^{\rm T})$, where $h$ is the thickness of the air gap, and $\epsilon _{\rm d}$ is the permittivity of the semi-infinite dielectric substrate. This dispersion relation entails that $k_{\rm sp}$ diverges as $h\to 0$. Moreover, the local SP mode exists when $-\epsilon_{\rm d}<\epsilon_{\rm m}^{\rm T}(\omega)<-1$, which agrees well with the
frequency range in Fig.~\ref{fig:wire_touches_substrate_resonances} where the local extinction cross section is large.

In the nonlocal HDM on the other hand, the dispersion relation changes into $\tanh ({k_{\rm sp}} h) =  - ({\epsilon _{\rm d}} + \epsilon _{\rm m}^{\rm T}+\epsilon _{\rm d}\Delta_{\rm L})/(1 + {\epsilon _{\rm d}}\epsilon _{\rm m}^{\rm T}+\Delta_{\rm L})$, where $\Delta_{\rm L}=k_{\rm sp}(\epsilon_{\rm m}^{\rm T}-1)/\kappa^{\rm L}$, and $-(\kappa^{\rm L})^2+k_{\rm sp}^2=(k^{\rm L})^2$. The nonlocal correction term $\Delta_{\rm L}$ regularizes the dispersion relation in the limit $h\to 0$, and thus makes the nonlocal extinction curves for a nanowire resting on a substrate completely different from the local one.

\begin{figure}[ht]
\includegraphics[width=0.45\textwidth]{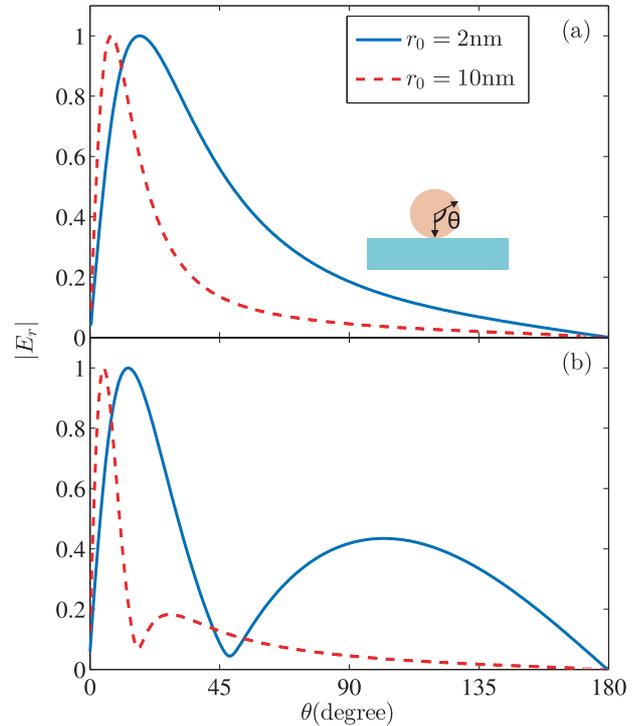}
\caption
{For the same nanowire resting on a substrate as in Fig.~\ref{fig:wire_touches_substrate_resonances}, the normal component of the electric field scaled with respect to its maximal value along the nanowire boundary, for two nanowire radii. Panel~(a): first-order nonlocal SP resonance mode; panel~(b): second-order nonlocal SP resonance mode.}\label{fig:wire_touches_substrate_radial_profile}
\end{figure}

In Fig.~\ref{fig:wire_touches_substrate_radial_profile}(a) and (b),
we plot for nonlocal response the normal component of the electric field along the nanowire boundary for the first- and second-order resonance modes observed in Fig.~\ref{fig:wire_touches_substrate_resonances}. Compared to Fig.~\ref{Fig:Enormaltheta} for a nanowire one nanometer away from the substrate, the field distribution in Fig.~\ref{fig:wire_touches_substrate_radial_profile} gained more weight on the substrate side. For  example for $r_0=10\,{\rm nm}$, the field amplitude peak in Fig.~\ref{Fig:Enormaltheta}(a) occurs near $\theta=20^{\rm o}$, and decreases to $\theta=8^{\rm o}$ in Fig.~\ref{fig:wire_touches_substrate_radial_profile}(a). The nanowire and the substrate increasingly influence each other as the nanowire approaches the substrate.

For the nanowire touching the substrate, Eq.~(\ref{effective_relative_nonlocal_blueshift}) becomes invalid since its derivation relies on the assumption that differences between local and nonlocal response are small. However, as we will see the trend described by Eq.~(\ref{effective_relative_nonlocal_blueshift}) that $\omega_{\rm res}^{\rm nloc}$ blueshifts as $r_0$ decreases still holds true, and thinking in terms of the effective parameters  $l_{\rm eff}$ and $\epsilon_{\rm b}^{\rm eff}$ is still useful.  The argument runs as follows.

Consider two plasmonic nanowires with radii $r_1>r_2$ and nonlocal SP resonance frequencies $\omega_{\rm res,1}^{\rm nloc}$ and $\omega_{\rm res,2}^{\rm nloc}$. Expanding the coordinate system isotropically, the nanowire with the radius $r_2$ could  equivalently be viewed as the nanowire with the
larger radius $r_1$ in combination with a nonlocal charge layer thickness $l_{\rm eff}$ that is increased by a factor of $r_1/r_2$. The increased charge layer results in a stronger field inside the nanowire proportional to $l_{\rm eff}$ as discussed in Sec.~\ref{sec:effective}. There will also be a weaker screening contribution from the substrate (i.e., the $\epsilon_{\rm b}^{\rm eff}$ is smaller for the smaller wire radius)  since the effective distance between the charge and the substrate is increased. These two effects make the resonance frequency $\omega_{\rm res,2}^{\rm nloc}$ for the smaller nanowire radius blueshifted with respect to $\omega_{\rm res,1}^{\rm nloc}$ also in the touching geometry. Incidentally, the latter effect agrees with Fig.~\ref{fig:leff_average_epsilon}(a1) where $\epsilon_{\rm b}^{\rm eff}$ also increases as a function of $r_{0}$.

\begin{figure}[ht]
\includegraphics[width=0.45\textwidth]{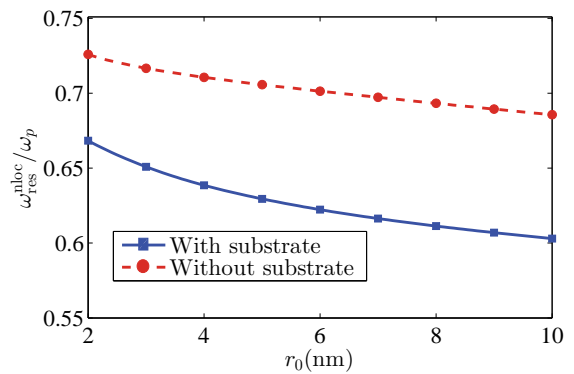}
\caption
{First-order resonance $\omega_{\rm res}^{\rm nloc}$ as a function of $r_0$ for  of a gold cylindrical nanowire resting on a semi-infinite dielectric substrate of index 1.5, in a free-space background. The case without a substrate is also shown. The data points are numerically exact values obtained with our nonlocal GSIM. The curves through the data points are guides to the eye. }\label{FIG:Comparison_of_blueshits_touching_vs_no_substrate}
\end{figure}

In Fig.~\ref{FIG:Comparison_of_blueshits_touching_vs_no_substrate}, we show  $\omega_{\rm res}^{\rm nloc}$ versus $r_0$ for the first-order nonlocal resonance. The nanowire-substrate touching geometry is compared to the case without the substrate. It is observed that $\omega_{\rm res}^{\rm nloc}$ blueshifts  for decreasing  wire radius, and more so with the substrate in place. We attribute this to the substrate-increased $l_{\rm eff}$ and $\epsilon_{\rm b}^{\rm eff}$.

\section{Discussion and conclusions}\label{Sec:discussion_and_conclusions}
In this paper, we generalized the local-response Green-function surface-integral method to a nonlocal version, where the nonlocal response is described by the hydrodynamical Drude model. The method developed here works for arbitrarily shaped nanowires in arbitrary inhomogeneous backgrounds.  The key insight that lead to our nonlocal GSIM is that an additional surface integral can be formulated that describes the Maxwell fields associated with the hydrodynamic pressure waves. Spill-out of free electrons is neglected, so their nonlocal response can be described in terms of the fields on the surfaces that confine them. Besides Maxwell's boundary conditions, there is an additional boundary condition that is easily derived once electron spill-out is neglected.

The GSIM has the advantage of being numerically light, but until now it was not clear how to apply even the known local-reponse GSIM to nanoparticles resting on substrates, surely a typical situation in experiments. We showed how to apply the GSIM in this case, by regularizing the singularities that only arise for such `touching geometries', by which we mean that the surface that is to be integrated over touches an interface. This regularization procedure works both for the local and for our nonlocal GSIM.  This makes the GSIM a more general-purpose numerical method, and we expect that this development will contribute to its popularity.

We expect the nonlocal GSIM also to become a method of choice when studying nonlocal response in complex geometries. Nonlocal response changes the charge distribution especially near the metal-dielectric interfaces, and it is only these interfaces that we need to discretize for the surface method. Thus nonlocal GSIM is computationally efficient and stays close to the action, so to say.

We first compared the nonlocal and local response of nanowires without substrates. We benchmarked the nonlocal GSIM against the analytical solution of the extinction of a cylindrical nanowire and found excellent agreement. We observed  the characteristic nonlocal blueshift of extinction resonances. However, our finding that the blueshift is linearly proportional to the angular momentum number of the surface-plasmon resonance is new, as far as we know. We also found an analytical derivation for this phenomenon, based on the fact that nonlocal response effectively pushes the surface-charge density inward into the plasmonic nanowire. It would be interesting for future studies to study the angular-momentum dependence of resonance frequency shifts when also allowing for spill-out of the free electrons.\cite{Teperik:2013}

For a nanowire as close as 1 nm to a dielectric substrate, we still can accurately account for the nonlocal blueshift of the resonances, using an effective theory. Besides the nanowire radius and the longitudinal wavevector, this involves an effective angular-momentum number and an effective background dielectric function. The nonlocal blueshift of a cylindrical nanowire is enhanced when close to a substrate. Our explanation can be summarized as follows: for an angular-momentum resonance of the nanowire, the substrate makes the charge distribution on the wire surface more inhomogeneous. The angular-momentum expansion of this charge distribution therefore involves higher angular-momentum numbers, and an effectively higher angular momentum can be defined with concomitant larger nonlocal blueshift. We also find that the substrate increases the effective background dielectric function. Both effects together accurately predict the enhanced nonlocal blueshift. We find that the effective background dielectric function varies by 20$\%$ when varying the nanowire radius from 2 nm to 18 nm. By contrast, the effective background dielectric function of nanospheres on a substrate was assumed to be independent of the sphere radius in Ref.~\onlinecite{Scholl:2013}.

We also calculated extinction spectra of nanowires resting on a dielectric surface. Pronounced differences are found between the local and the nonlocal theory, so that our effective theory for nonlocal blueshifts does not work here. Similar large differences have been predicted before for plasmonic dimers structures (two spheres,\cite{Fuchs:1987} or two wires\cite{FernandezDominguez:2012}). Here we show that a single plasmonic nanowire on a dielectric substrate is already enough to observe considerable differences between local and nonlocal response. It may also be the preferred experimental structure to study strong nonlocal effects, since quantum tunneling as for plasmonic dimers is less of a complication.

In this paper we focused on extinction cross sections of nanowires, but also waveguiding, electron energy-loss spectroscopy, and other observables could be calculated using our method. Moreover, the method does not only work for the nanowire structures considered here. We are presently generalizing our nonlocal GSIM to truly three-dimensional geometries, where advantages of surface integral methods are even more pronounced. The general idea is the same, namely to add to the known surface integrals a 3D version of the surface integral for the longitudinal field.

\section*{Acknowledgments.}
This work was financially supported by an H.~C.~{\O}rsted Fellowship (W.Y.). W.Y. would like to thank Thomas S{\o}ndergaard for introducing him to the Green-function surface-integral method. The Center for Nanostructured Graphene is sponsored by the Danish National Research Foundation, Project DNRF58.

\appendix

\section{Derivations of Surface Integrals }\label{App:surface_integrals}
\subsection{Derivation of Eqs.~(\ref{trasieq0}) and (\ref{logsieq})}\label{sec:GSIM_local_eqs_metal}
Consider the area integral
 \begin{equation}
 \int_{{A_{i}}} \mbox{d}x\mbox{d}y\,\left[ E_{zi}(\boldsymbol \rho)\nabla_{\boldsymbol\rho}^2{g_{i}^{\rm T}}(\boldsymbol \rho,\boldsymbol \rho') - g_{i}^{\rm T}(\boldsymbol \rho,\boldsymbol \rho ')\nabla_{\boldsymbol\rho}^2 E_{zi}(\boldsymbol \rho)\right],
\label{eq1}
\end{equation}
where $x,y\in A_{i}$. From the definition of the scalar Green function $g_{i}^{\rm T}$ in Sec.~\ref{sec:GSIM_local}, it follows that Eq.~(\ref{eq1}) is identical to $E_{zi}(\boldsymbol\rho')$. Using the identity $\phi\nabla^2\psi=\boldsymbol\nabla\cdot\phi\boldsymbol\nabla\psi-\boldsymbol\nabla\phi\cdot\boldsymbol\nabla\psi$ and Gauss's integral theorem, Eq.~(\ref{eq1}) can be rewritten as
\begin{eqnarray}
\oint_{S_i} \mbox{d}{\boldsymbol\rho} &&[  E_{zi}(\boldsymbol\rho){\hat n_{i}(\boldsymbol\rho)} \cdot \boldsymbol\nabla_{\boldsymbol\rho} {g_{i}^{\rm T}}(\boldsymbol\rho,\boldsymbol\rho') \nonumber \\
&-& {g_{i}^{\rm T}}(\boldsymbol \rho,\boldsymbol\rho' ){\hat n_{i}(\boldsymbol\rho)} \cdot \boldsymbol\nabla_{\boldsymbol\rho} E_{zi}(\boldsymbol \rho)].
\label{eq2}
\end{eqnarray}
By equating Eq.~(\ref{eq2}) with $E_{zi}$, and by using the reciprocity property that $g_{i}^{\rm T}(\boldsymbol \rho,\boldsymbol \rho')=g_{i}^{\rm T}(\boldsymbol \rho',\boldsymbol \rho)$, one arrives at the surface integral for the metal domains Eq.~(\ref{trasieq0}). The additional surface integral Eq.~(\ref{logsieq}) for nonlocal response in the metal can be derived analogously.

\subsection{Derivation of Eq.~(\ref{trabaeq})}\label{sec:GSIM_local_eqs_background}
When the background is spatially inhomogeneous, it is difficult to follow the same routine as used in Sec.~\ref{sec:GSIM_local_eqs_metal} above to derive the surface integrals for the dielectric side of the metal-dielectric boundaries. In Ref.~\onlinecite{Jung:2008}, the surface integrals for the specific inhomogeneous background with the planar interface are derived by matching the boundary conditions of the Green-function at the interface. Here, we employ an alternative approach based on the surface equivalence theorem to derive the surface integrals for arbitrary backgrounds.\cite{Kong:2005} Denoting the actual field distribution as $\bigl\{\mathbf E_{\rm b}(\boldsymbol \rho),\,\mathbf H_{\rm b}(\boldsymbol \rho)\bigl\}$ for $\boldsymbol \rho \in B$, and $\bigl\{\mathbf E_{\rm i}(\boldsymbol \rho),\,\mathbf H_{\rm i}(\boldsymbol \rho)\bigl\}$ for $\boldsymbol \rho \in A$. Then, consider a virtual field distribution with $\bigl\{\mathbf E_{\rm i}(\boldsymbol \rho),\,\mathbf H_{\rm i}(\boldsymbol \rho)\bigl\}$ replaced by $\bigl\{{\bm 0},\,{\bm 0}\bigl\}$. The existence of such a virtual field distribution requires a set of surface currents\cite{Kong:2005}
\begin{subequations}
\begin{eqnarray}
&\,&\widetilde{\mathbf J}_e(\boldsymbol \rho)= \hat n(\boldsymbol \rho)\times \mathbf H_b(\boldsymbol \rho),\\
&\,&\widetilde{\mathbf M}_e(\boldsymbol \rho)= -\hat n(\boldsymbol \rho)\times \mathbf E_b(\boldsymbol \rho),
\end{eqnarray}
\label{eqscurrent}
\end{subequations}
existing only on the metal-dielectric boundary S, and where $\widetilde{\mathbf J}_e$ and $\widetilde{\mathbf M}_e$ represent the surface electric and magnetic currents, respectively. The surface currents fix the unphysical field discontinuities across the boundary. The virtual fields are equivalently a result of the fields radiated by $\widetilde{\mathbf J}_e$, $\widetilde{\mathbf M}_e$, and also $\widetilde{\mathbf J}_b$, i.e.,
\begin{subequations}
\begin{eqnarray}
{\mathbf E_b}(\boldsymbol\rho )&=&{\mathbf E_b^{\rm inc}}(\boldsymbol\rho )+\int \mbox{d}\boldsymbol\rho' \,{\bfsfG }_e(\boldsymbol\rho,\boldsymbol\rho ') \cdot {i\omega {\mu _0}{ {\widetilde{\mathbf J}_e}(\boldsymbol\rho')}},\nonumber\\
&+&\int \mbox{d}\boldsymbol\rho' \,{\bfsfG }_e(\boldsymbol\rho,\boldsymbol\rho ') \cdot\left[\left( - i{k_z}\hat z \times-{\boldsymbol\nabla _{\boldsymbol\rho'}}\right) \times
{\widetilde{\mathbf M}_e(\boldsymbol\rho')}\right]\nonumber\\
\\
{\mathbf H_b}(\boldsymbol\rho )&=&{\mathbf H_b^{\rm inc}}(\boldsymbol\rho )+\int \mbox{d}\boldsymbol\rho' \,{\bfsfG }_m(\boldsymbol\rho ,\boldsymbol\rho ') \cdot i\omega {\epsilon _0}{\widetilde{\mathbf M}_e}(\boldsymbol\rho ')\nonumber\\
&+&\int \mbox{d}\boldsymbol\rho' \,{\bfsfG }_m(\boldsymbol\rho ,\boldsymbol\rho ')\cdot\left[\left( i{k_z}\hat z \times+{\boldsymbol\nabla _{\boldsymbol\rho'}}\right) \times
{\widetilde{\mathbf J}_e(\boldsymbol\rho')}\right].\nonumber\\
\end{eqnarray}
\label{fieldb1}
\end{subequations}
${\mathbf E_b^{\rm inc}}$ and ${\mathbf H_b^{\rm inc}}$ represent the incident fields from $\widetilde{\mathbf J}_b$. Furthermore,  ${\bfsfG }_{e}$ and ${\bfsfG }_{m}$ represent the background electric and magnetic dyadic Green functions, defined by
\begin{subequations}
\begin{eqnarray}
&\,& \left[\boldsymbol\nabla_{k_z} \times \boldsymbol\nabla_{k_z}  \times- k_0^2\epsilon_b(\boldsymbol\rho)\right] {\bfsfG_e}\left( {\boldsymbol\rho ,\boldsymbol\rho '} \right)  = \bfsfI \delta (\boldsymbol\rho  - \boldsymbol\rho '),\nonumber\\
\\
&\,&\left[\boldsymbol\nabla_{k_z}  \times\frac{1}{\epsilon_b(\boldsymbol\rho)}\boldsymbol\nabla_{k_z}  \times -  k_0^2\right] {\bfsfG_m}\left( {\boldsymbol\rho ,\boldsymbol\rho '} \right) = \bfsfI \delta (\boldsymbol\rho  - \boldsymbol\rho '),\nonumber\\
\end{eqnarray}
\label{GEM}
\end{subequations}
where $\boldsymbol\nabla_{k_z}  =\left( {{\boldsymbol\nabla _{\boldsymbol\rho} } + i{k_z}\hat z} \right)$ and  $\bfsfI$ is the $3\times 3$ unit matrix. Extracting the $z$-component of $\mathbf E_b$ and $\mathbf H_b$ and taking the expressions of the surface currents into Eq.~(\ref{fieldb1}), we then derive the surface integrals of Eq.~(\ref{trabaeq}) for the fields on the dielectric side of the metal-dielectric boundaries, valid for arbitrary spatial inhomogeneity $\epsilon_{b}({\bm \rho})$ of the background.

Until now we have assumed that the metal nanowires are surrounded by a dielectric background, but let us discuss briefly how to describe the situation that there is also metal in the background, for example a metal substrate for plasmonic nanoparticles as in the recent experiments by Oulton {\em et al.}~\cite{Oulton:2009} and by Cirac{\` i} {\em et al.}~\cite{Ciraci:2012b} If we neglect possible nonlocal response of the metal in the background, then the optical response of the metal can also be described by the spatially inhomogeneous but local dielectric function $\epsilon_b({\boldsymbol\rho})$, so the above formalism can be applied.

Alternatively, if one would like to describe the metal in the background also by the hydrodynamical Drude model, then we can do this by taking $\epsilon_b$ in the dynamic Green functions $\bfsfG_e$ and $\bfsfG_m$ of Eq.~(\ref{GEM}) to be a nonlocal operator defined by Eqs.~(\ref{hdm}) and~(\ref{lortz}). In other words, we absorb possible nonlocal effects of the background into the dynamic Green functions $\bfsfG_e$ and $\bfsfG_m$. Then, the nonlocal response is contained in the surface-integral coefficients of Eq.~(\ref{bcoef1}). This differs from our treatment of the plasmonic nanowires, where we decomposed the fields into the longitudinal and transverse parts. Absorbing any plasmonic nonlocal response of the background into the background Green tensor is not just a formal trick. For example, for inhomogenous backgrounds of a plasmonic slab substrate in free space, the corresponding surface-integral kernels for the nonlocal GSIM  can be found in Appendix~\ref{Sec:slab_background}.


\section{Integration kernels in Eq.~(\ref{bcoef1}) for a layered substrate}\label{Sec:slab_background}
Here we consider inhomogeneous backgrounds that can be described as substrates that are arbitrary planar multilayer systems  in free space. We choose a convenient coordinate system such that the substrate of thickness $t$ is located at $0<x<t$, with the nanowires in the region $x<0$. A semi-infinite substrate would correspond to $t=\infty$. The substrate consists of dielectric or  metal slabs or a combination thereof. Any metallic layers can either be described with local or with nonlocal response. For all those cases, we present the integration kernels for the surface integral for the fields in the region $x<0$ outside of the nanowires. This can be done because in the region $x<0$ the background Green tensors $\bfsfG_{e,m}({\bm \rho},{\bm \rho'})$ and hence the kernels in Eq.~(\ref{bcoef1}) can be expressed in terms of the substrate reflection coefficients at $x=0$; only the values of these reflection coefficients are different for different metal-dielectric multilayers systems, and also different if the metals are described with local or nonlocal response. For the actual calculation of these reflection coefficients, we refer to textbooks, for example Ref.~\onlinecite{Chew:1990}; for the Green function in layered geometries to Ref.~\onlinecite{Tomas:1995}, and for wave propagation in multilayer systems with nonlocal response to Refs.~\onlinecite{Mochan:1987} and~\onlinecite{Yan:2012}.

We split the dyadic Green function ${\bfsfG }_{e,m}$ into a homogenous and a scattering part, i.e. we write ${\bfsfG }_{e,m}={\bfsfG }_{e,m}^0+{\bfsfG }_{e,m}^s$.\cite{Tomas:1995} Accordingly, the integration kernels in Eq.~(\ref{bcoef1}) are split into homogenous and scattering parts, for example $e_{b}^{0}=e_{b}^{00}+e_{b}^{0s}$. The homogeneous parts of the integration kernels are discussed in Sec.~\ref{sec:GSIM_local}, while the scattering parts can be derived as
\begin{widetext}
\begin{subequations}
\begin{eqnarray}
e_{b}^{0s}(\boldsymbol\rho,\boldsymbol\rho')&=&\frac{i}{{4\pi }}\int \mbox{d}{{k_y}} \frac{1}{{{k_{x}}}}\exp(i\psi)i\mathbf k_{\rho}\cdot\hat n(\boldsymbol\rho')
\left[-R_{\rm TE}(k_{\parallel})\frac{k_y^2k_0^2}{k_{\rho}^2k_{\parallel}^2}+R_{\rm TM}(k_{\parallel})\frac{k_x^2k_z^2}{k_{\rho}^2k_{\parallel}^2}\right],\label{eb0s} \\
e_{b}^{1s}(\boldsymbol\rho,\boldsymbol\rho')&=&\frac{i}{{4\pi }}\int \mbox{d}{{k_y}} \frac{1}{{{k_{x}}}}\exp (i\psi)
\left[-R_{\rm TE}(k_{\parallel})\frac{k_y^2k_0^2}{k_{\rho}^2k_{\parallel}^2}+R_{\rm TM}(k_{\parallel})\frac{k_x^2k_z^2}{k_{\rho }^2k_{\parallel}^2}\right],\\
f_{b}^{0s}(\boldsymbol\rho,\boldsymbol\rho')&=&\frac{i}{{4\pi }}\int \mbox{d}{{k_y}} \frac{\omega\mu_0}{{{k_{x}}}}\exp (i\psi)i\mathbf k_{\rho}\cdot\hat n(\boldsymbol\rho')\frac{k_{x}k_yk_z}{k_{\rho}^2k_{\parallel}^2}
\left[R_{\rm TE}(k_{\parallel})+R_{\rm TM}(k_{\parallel})\right],\\
f_{b}^{1s}(\boldsymbol\rho,\boldsymbol\rho')&=&\frac{i}{{4\pi }}\int \mbox{d}{{k_y}} \frac{\omega\mu_0}{{{k_{x}}}}\exp (i\psi)\frac{k_{x}k_yk_z}{k_{\rho }^2k_{\parallel}^2}
\left[R_{\rm TE}(k_{\parallel})+R_{\rm TM}(k_{\parallel})\right],\\
h_{b}^{0s}(\boldsymbol\rho,\boldsymbol\rho')&=&\frac{-i}{{4\pi }}\int \mbox{d}{{k_y}} \frac{\omega\epsilon_0}{{{k_{x}}}}\exp(i\psi)i\mathbf k_{\rho}\cdot\hat n(\boldsymbol\rho')\frac{k_{x}k_yk_z}{k_{\rho }^2k_{\parallel}^2}
\left[R_{\rm TE}(k_{\parallel})+R_{\rm TM}(k_{\parallel})\right],\\
h_{b}^{1s}(\boldsymbol\rho,\boldsymbol\rho')&=&\frac{-i}{{4\pi }}\int \mbox{d}{{k_y}}\frac{\omega\epsilon_0}{{{k_{x}}}}\exp(i\psi)\frac{k_{x}k_yk_z}{k_{\rho }^2k_{\parallel}^2}
\left[R_{\rm TE}(k_{\parallel})+R_{\rm TM}(k_{\parallel})\right],\\
m_{b}^{0s}(\boldsymbol\rho,\boldsymbol\rho')&=&\frac{i}{{4\pi }}\int \mbox{d}{{k_y}} \frac{1}{{{k_{x}}}}\exp(i\psi)i\mathbf k_{\rho}\cdot\hat n(\boldsymbol\rho')
\left[R_{\rm TE}(k_{\parallel})\frac{k_x^2k_z^2}{k_{\rho}^2k_{\parallel}^2}-R_{\rm TM}(k_{\parallel})\frac{k_y^2k_0^2}{k_{\rho}^2k_{\parallel}^2}\right], \\
m_{b}^{1s}(\boldsymbol\rho,\boldsymbol\rho')&=&\frac{i}{{4\pi }}\int \mbox{d}{{k_y}} \frac{1}{{{k_{x}}}}\exp(i\psi)
\left[R_{\rm TE}(k_{\parallel})\frac{k_x^2k_z^2}{k_{\rho}^2k_{\parallel}^2}-R_{\rm TM}(k_{\parallel})\frac{k_y^2k_0^2}{k_{\rho}^2k_{\parallel}^2}\right],
\end{eqnarray}
\label{ibsicoef}
\end{subequations}
\end{widetext}
where $\boldsymbol \rho = (x,y)$ is in the region $x<0$, $\psi={k_y}(y - y')-k_{x}(x+x')$, $k_0=\omega/c$, $k_x^2+k_y^2+k_z^2=k_0^2$, $k_\rho^2=k_0^2-k_z^2$, $k_y^2+k_z^2=k_{\parallel}^2$; $R_{\rm TE}$ and $R_{\rm TM}$ represent the reflection coefficients at $x=0$ of the multilayer substrate for TE and TM polarized plane waves, respectively.

\section{Derivation of Eq.~(\ref{eqeffl})}\label{Sec:l_eff_def}
Here we derive the effective angular momentum number $l_{\rm eff}$ for the SP mode supported by a cylindrically shaped plasmonic nanowire in an inhomogeneous background. This is a key parameter in our explanation of nonlocal blueshifts  of nanowires in arbitrary dielectric backgrounds, especially in  Eq.~(\ref{effective_relative_nonlocal_blueshift}). The arguments used here are similar to those developed in Sec.~\ref{sec:effective}.

Let us first consider a cylindrical nanowire with local response, in an inhomogeneous background. Define $\alpha_{\rm m}$ as the surface charge at the nanowire boundary $r=r_0$ of the SP mode of  the wire.  The surface charge can be decomposed into
\begin{equation}
\alpha_{\rm m}={\sum \limits_{l \ne 0}}\alpha_{\rm ml}\exp(il\phi)\delta(r-r_0),
\end{equation}
where $l$ is the angular momentum number of the cylindrical harmonics that ranges from $-\infty$ to $\infty$; The term $l=0 $ is excluded from the summation, as it does not contribute to the surface charge. The surface charge $\alpha_{\rm m}$ is a source that  generates electric fields $\mathbf E$ inside the nanowire given by
\begin{eqnarray}
\mathbf E=\frac{1}{2\epsilon_0}{\sum \limits_{l \ne 0}} \alpha_{\rm ml}\left(\frac{r}{r_0}\right)^{l-1}\exp(il\phi)(\hat r+i\hat \phi).
\end{eqnarray}

Second, we consider the same structure, but now we describe the nanowire with nonlocal response. The corresponding surface charge $\alpha_{\rm m}^\prime$ will now effectively be moved inwards into the nanowire, to $r_0^\prime=r_0-1/k^{\rm L}$, and is expressed as
\begin{equation}
\alpha_{\rm m}^\prime={\sum \limits_{l \ne 0}} \alpha_{\rm ml}\frac{r_0}{r_0^\prime}\exp(il\phi)\delta(r-r_0^\prime).
\end{equation}
In the region inside $r_0^\prime$, denoted by $A_{\rm m}^\prime$, the charge density $\alpha_{\rm m}^\prime$ generates the electric field
\begin{eqnarray}
\mathbf E^\prime=\frac{1}{2\epsilon_0}{\sum \limits_{l \ne 0}} \alpha_{\rm ml}\frac{r_0}{r_0^\prime}\left(\frac{r}{r_0^\prime}\right)^{l-1}\exp(il\phi)(\hat r+i\hat \phi).
\end{eqnarray}
Consequently, the electric field in the area $A_{\rm m}^\prime$ on average is enhanced by a factor
\begin{eqnarray}
F&=&\sqrt{\frac{\int_{A_{\rm m}^\prime}d\mathbf r |\mathbf E^\prime|^2}{\int_{A_{\rm m}^\prime}d\mathbf r |\mathbf E|^2}}\nonumber\\
&\approx&\left(\frac{r_0}{r_0^\prime}\right)^{\frac{{\sum\limits_{l \ne 0}}|\alpha_{\rm ml}|^2}{{\sum\limits_{l \ne 0}}|\alpha_{\rm ml}|^2l^{-1}}}.
\label{eqfactor}
\end{eqnarray}
Now in a homogenous background, where the angular-momentum number corresponding to the order of the cylindrical  harmonics is well defined, the factor $F$ is equal to $(r_0/r_0^{\prime})^l$. In our effective description we now identify Eq.~(\ref{eqfactor}) with $F=(r_0/r_0^{\prime})^{l_{\rm eff}}$, which allows us to extract the effective angular momentum number $l_{\rm eff}$ as
\begin{equation}
l_{\rm eff}^{-1}={\frac{{\sum\limits_{l \ne 0}}|\alpha_{\rm ml}|^2l^{-1}}{{\sum\limits_{l \ne 0}}|\alpha_{\rm ml}|^2}},
\end{equation}
which is Eq.~(\ref{eqeffl}) of the main text.

The effective angular momentum $l_{\rm eff}$ so defined is only determined by the surface distribution of the free charges and the associated electric field in the plasmonic nanowire. In other words, for $l_{\rm eff}$ we do not consider the electric field that can be associated with  screened charges in the inhomogeneous background. Those screened charges influence the other effective parameter, namely the effective background permittivity $\epsilon_{\rm b}^{\rm eff}$ in Eq.~(\ref{effective_relative_nonlocal_blueshift}).


\begin{thebibliography}{10}

\bibitem{Maier:2007}
S.~A. Maier, {\em Plasmonics: Fundamentals and Applications} (Springer, New
  York, 2007).

\bibitem{Gramotnev:2010}
D.~K. Gramotnev and S.~I. Bozhevolnyi, {\em Plasmonics beyond the diffraction
  limit}, Nat. Photonics {\bf 4},  83  (2010).

\bibitem{Nelayah:2007}
J. Nelayah, M. Kociak, O. Stephan, F.~J. {Garc\'{i}a de Abajo}, M. Tence, L.
  Henrard, D. Taverna, I. Pastoriza-Santos, L.~M. Liz-Marzan, and C. Colliex,
  {\em Mapping surface plasmons on a single metallic nanoparticle}, Nat. Phys.
  {\bf 3},  348  (2007).

\bibitem{Nicoletti:2011}
O. Nicoletti, M. Wubs, N.~A. Mortensen, W. Sigle, P.~A. van Aken, and P.~A.
  Midgley, {\em Surface plasmon modes of a single silver nanorod: an electron
  energy loss study}, Opt. Express {\bf 19},  15371  (2011).

\bibitem{Scholl:2012}
J.~A. Scholl, A.~L. Koh, and J.~A. Dionne, {\em Quantum plasmon resonances of
  individual metallic nanoparticles}, Nature {\bf 483},  421  (2012).

\bibitem{Ciraci:2012b}
C. Cirac{\` i}, R.~T. Hill, J.~J. Mock, Y. Urzhumov, A.~I. Fern{\'
  a}ndez-Dom{\' i}nguez, S.~A. Maier, J.~B. Pendry, A. Chilkoti, and D.~R.
  Smith, {\em Probing the ultimate limits of plasmonic enhancement}, Science
  {\bf 337},  1072  (2012).

\bibitem{Kern:2012}
J. Kern, S. Grossmann, N.~V. Tarakina, T. H\"ackel, M. Emmerling, M. Kamp,
  J.-S. Huang, P. Biagioni, J.~C. Prangsma, and B. Hecht, {\em Atomic-scale
  confinement of resonant optical fields}, Nano Lett. {\bf 12},  5504  (2012).

\bibitem{Savage:2012}
K.~J. Savage, M.~M. Hawkeye, R. Esteban, A.~G. Borisov, J. Aizpurua, and J.~J.
  Baumberg, {\em Revealing the quantum regime in tunnelling plasmonics}, Nature
  {\bf 491},  574  (2012).

\bibitem{Scholl:2013}
J. Scholl, A. Garcia-Etxarri, A.~L. Koh, and J.~A. Dionne, {\em Observation of
  quantum tunneling between two plasmonic nanoparticles}, Nano Lett. {\bf 13},
  564  (2013).

\bibitem{Teperik:2013}
T.~V. Teperik, P. Nordlander, J. Aizpurua, and A.~G. Borisov, {\em Robust
  subnanometric plasmon ruler by rescaling of the nonlocal optical response},
  Phys. Rev. Lett. {\bf 110},  263901  (2013).

\bibitem{Zuloaga:2009}
J. J.~Zuloaga, E. Prodan, and P. Nordlander, {\em Quantum description of the
  plasmon resonances of a nanoparticle dimer}, Nano Lett. {\bf 9},  887–891
  (2009).

\bibitem{Toscano:2012a}
G. Toscano, S. Raza, A.-P. Jauho, M. Wubs, and N.~A. Mortensen, {\em Modified
  field enhancement and extinction in plasmonic nanowire dimers due to nonlocal
  response}, {Opt. Express} {\bf {13}},  4176  ({2012}).

\bibitem{Dong:2012}
T. Dong, X. Ma, and R. Mittra, {\em Optical response in subnanometer gaps due
  to nonlocal response and quantum tunneling}, Appl. Phys. Lett. {\bf 101},
  233111  (2012).

\bibitem{Bloch:1933}
F. Bloch, {\em Bremsverm\text{\"{o}}gen von Atomen mit mehreren Elektronen}, Z.
  Phys. {\bf 81},  363  (1933).

\bibitem{Ying:1974}
S.~C. Ying, {\em Hydrodynamic response of inhomogenous metallic systems}, Nuovo
  Cimento B {\bf 23},  270  (1974).

\bibitem{Eguiluz:1976}
A. Eguiluz and J.~J. Quinn, {\em Hydrodynamic model for surface plasmons in
  metals and degenerate semiconductors}, Phys. Rev. B {\bf 14},  1347  (1976).

\bibitem{Boardman:1982}
A.~D. Boardman, {\em Electromagnetic Surface Modes} (John Wiley and Sons,
  Chichester, 1982).

\bibitem{Raza:2011}
S. Raza, G. Toscano, A.-P. Jauho, M. Wubs, and N.~A. Mortensen, {\em Unusual
  resonances in nanoplasmonic structures due to nonlocal response}, {Phys. Rev.
  B} {\bf {84}},  121412(R)  ({2011}).

\bibitem{Ruppin:1992}
R. Ruppin, {\em Optical absorption by a small sphere above a substrate with
  inclusion of nonlocal effects}, Phys. Rev. B {\bf 45},  11209  (1992).

\bibitem{Ruppin:1973}
R. Ruppin, {\em Optical properties of a plasma sphere}, Phys. Rev. Lett. {\bf
  31},  1434  (1973).

\bibitem{Ruppin:2001}
R. Ruppin, {\em Extinction properties of thin metallic nanowires}, Opt. Commun.
  {\bf 190},  205  (2001).

\bibitem{Fuchs:1987}
R. Fuchs and F. Claro, {\em Multipolar response of small metallic spheres:
  nonlocal theory}, Phys. Rev. B {\bf 35},  3722  (1987).

\bibitem{FernandezDominguez:2012}
A.~I. Fern{\' a}ndez-Dom{\' i}nguez, A. Wiener, F.~J. Garc{\' i}a-Vidal, S.~A.
  Maier, and J.~B. Pendry, {\em Transformation-optics description of nonlocal
  effects in plasmonic nanostructures}, Phys. Rev. Lett. {\bf 108},  106802
  (2012).

\bibitem{Marier:2012}
A. Wiener, A.~I. Fern{\' a}ndez-Dom{\' i}nguez, A.~P. Horsfield, J.~B. Pendry,
  and S.~A. Maier, {\em Nonlocal effects in the nanofocusing performance of
  plasmonic tips}, Nano Lett. {\bf 12},  3308  ({2012}).

\bibitem{Raza:2012a}
S. Raza, G. Toscano, A.-P. Jauho, N.~A. Mortensen, and M. Wubs, {\em
  Refractive-index sensing with ultrathin plasmonic nanotubes}, Plasmonics {\bf
  8},  193  (2013).

\bibitem{Toscano:2012b}
G. Toscano, S. Raza, S. Xiao, M. Wubs, A.-P. Jauho, S.~I. Bozhevolnyi, and
  N.~A. Mortensen, {\em Surface-enhanced Raman spectroscopy (SERS): nonlocal
  limitations}, {Opt. Lett.} {\bf 37},  2538  (2012).

\bibitem{Mochan:1987}
W.~L. Moch\'an, M. Castillo-Mussot, and R.~G. Barrera, {\em {Effect of plasma
  waves on the optical properties of metal-insulator superlattices}}, {Phys.
  Rev. B} {\bf {35}},  1088   ({1987}).

\bibitem{Abajo:2008}
F.~J. Garc{\' i}a~de Abajo, {\em {Nonlocal effects in the plasmons of strongly
  interacting nanoparticles, dimers, and waveguides}}, {J. Phys. Chem. C} {\bf
  {112}},  17983  ({2008}).

\bibitem{David:2011}
C. David and F.~J. Garc{\' i}a~de Abajo, {\em {Spatial nonlocality in the
  optical response of metal nanoparticles}}, {J. Phys. Chem. C} {\bf {115}},
  19470  ({2011}).

\bibitem{Sipe:1980}
J.~E. Sipe, V.~C.~Y. So, M. Fukui, and G.~I. Stegeman, {\em Analysis of
  second-harmonic generation at metal surfaces}, Phys. Rev. B {\bf 21},  4389
  (1980).

\bibitem{Ciraci:2012a}
C. Cirac{\` i}, E. Poutrina, M. Scalora, and D.~R. Smith, {\em Origin of
  second-harmonic generation enhancement in optical split-ring resonators},
  Phys. Rev. B {\bf 85},  201403(R)  (2012).

\bibitem{Yee:1966}
K.~S. Yee, {\em Numerical solution of initial boundary value problems involving
  Maxwell's equations in isotropic media}, IEEE Trans. Antennas Propag. {\bf
  14},  302  (1966).

\bibitem{Taflove:2005}
A. Taflove and S.~C. Hagness, {\em Computational Electrodynamics: The
  Finite-Difference Time-Domain Method} (Artech House, Boston/London, 2005).

\bibitem{Jin:2002}
J.~M. Jin, {\em The Finite Element Method in Electromagnetics} (Wiley, New
  York, 2002).

\bibitem{Abajo:1998}
F.~J. Garc{\' i}a~de Abajo and A. Howie, {\em Relativistic electron energy loss
  and electron-induced photon emission in inhomogeneous dielectrics}, Phys.
  Rev. Lett. {\bf 80},  5180  (1998).

\bibitem{Abajo:2002}
F.~J. Garc{\' i}a~de Abajo and A. Howie, {\em Retarded field calculation of
  electron energy loss in inhomogenous dielectrics}, Phys. Rev. B {\bf 65},
  115418  (2002).

\bibitem{Abajo:2006}
I. Romero, J. Aizpurua, G.~W. Bryant, and F.~J. Garc{\' i}a~de Abajo, {\em
  Plasmons in nearly touching metallic nanoparticles: singular response in the
  limit of touching dimers}, Opt. Express {\bf 14},  9988–  (2006).

\bibitem{Thomas:2007}
T. S{\o}ndergaard and S. Bozhevolnyi, {\em Slow-plasmon resonant
  nanostructures: Scattering and field enhancements}, Phys. Rev. B {\bf 75},
  073402  (2007).

\bibitem{Jung:2008}
J. Jung and T. S{\o}ndergaard, {\em {Green's} function surface integral
  equation method for theoretical analysis of scatterers close to a metal
  interface}, Phys. Rev. B {\bf 77},  245310  (2008).

\bibitem{Kern:2009}
A.~M. Kern and O.~J.~F. Martin, {\em Surface integral formulation for 3D
  simulations of plasmonic and high permittivity nanostructures}, J. Opt. Soc.
  Am. A {\bf 26},  732  (2009).

\bibitem{Gallinet:2010}
B. Gallinet, A.~M. Kern, and O.~J.~F. Martin, {\em Accurate and versatile
  modeling of electromagntic scattering on periodic nanostructures with a
  surface integral approach}, J. Opt. Soc. Am. A {\bf 65},  115418  (2010).

\bibitem{Hiremath:2012}
K.~R. Hiremath, L. Zschiedrich, and F. Schmidt, {\em Numerical solution of
  nonlocal hydrodynamic Drude model for arbitrary shaped nano-plasmonic
  structures using N{\'e}d{\'e}lec finite elements}, J. Comp. Phys. {\bf 231},
  5890 – 5896  (2012).

\bibitem{Huang:2013}
Q. Huang, F. Bao, and S. He, {\em Nonlocal effects in a hybrid plasmonic
  waveguide for nanoscale confinement}, Opt. Express {\bf 21},  1430  (2013).

\bibitem{Toscano:2012c}
G. Toscano, S. Raza, W. Yan, C. Jeppesen, S. Xiao, M. Wubs, A.-P. Jauho, S.~I.
  Bozhevolnyi, and N.~A. Mortensen, {\em Nonlocal response in plasmonic
  waveguiding with extreme light confinement}, Nanophotonics {\bf 2},  161
  (2013).

\bibitem{Kong:2005}
J.~A. Kong, {\em Electromagnetic Wave Theory} (EMW Publishing, Cambridge MA,
  2008).

\bibitem{Barton:1979}
G. Barton, {\em Some surface effects in the hydrodynamic model of metals}, Rep.
  Prog. Phys. {\bf 42},  963  (1979).

\bibitem{Yan:2012}
W. Yan, M. Wubs, and N.~A. Mortensen, {\em Hyperbolic metamaterials: nonlocal
  response regularizes broadband singularity}, Phys. Rev. B {\bf 86},  205429
  (2012).

\bibitem{Tomas:1995}
M.~S. Toma\v{s}, {\em Green function for multilayers: Light scattering in
  planar cavities}, {Phys. Rev. A} {\bf {51}},  2545  ({1995}).

\bibitem{Jewsbury:1981a}
P. Jewsbury, {\em Electrodynamic boundary conditions at metal interfaces}, J.
  Phys. F: Met. Phys. {\bf 11},  195  (1981).

\bibitem{Ford:1984}
G. Ford and W. Weber, {\em Electromagnetic interactions of molecules with metal
  surfaces}, Phys. Rep. {\bf 113},  195  (1984).

\bibitem{Chew:1990}
W.~C. Chew, {\em Waves and Fields in Inhomogeneous Media} (Van Nostrand
  Reinhold, New York, 1990).

\bibitem{Raza:2012b}
S. Raza, N. Stenger, S. Kadkhodazadeh, S.~V. Fischer, N. Kostesha, A.-P. Jauho,
  A. Burrows, M. Wubs, and N.~A. Mortensen, {\em Blueshift of the surface
  plasmon resonance in silver nanoparticles studied with EELS}, Nanophotonics
  {\bf 2},  131  (2013).

\bibitem{Boardman:1977}
A.~D. Boardman and B.~V. Paranjape, {\em The optical surface modes of metal
  spheres}, J. Phys. F: Met. Phys. {\bf 7},  1935  (1977).

\bibitem{Jung:2011}
J. Jung and T.~G. Pedersen, {\em Exact polarizability and plasmon resonances of
  partly buried nanowires}, Opt. Express {\bf 19},  22775  (2011).

\bibitem{Zhang:2012}
S. Zhang and H. Xu, {\em Optimizing substrate-mediated plasmon coupling toward
  high-performance plasmonic nanowire waveguides}, ACS Nano {\bf 6},  8128–8135
   (2012).

\bibitem{Kottmann:2001}
J.~P. Kottmann and O.~J.~F. Martin, {\em Plasmon resonant coupling in metallic
  nanowires}, Opt. Express {\bf 8},  665  (2001).

\bibitem{Nordlander:2003}
E. Prodan, C. Radloff, N.~J. Halas, and P. Nordlander, {\em A hybridization
  model for the plasmon response of complex nanostructures}, Science {\bf 302},
   419  (2003).

\bibitem{Nordlander:2005}
D.~W. Brandl, C. Oubre, and P. Nordlander, {\em Plasmon hybridization in
  nanoshell dimers}, J. Chem. Phys. {\bf 123},  024701  (2005).

\bibitem{Davis:2010}
T.~J. Davis, D.~E. Gom{\' e}z, and K.~C. Vernon, {\em Simple model for the
  hybridization of surface plasmon resonances in metallic nanoparticles}, Nano.
  Lett. {\bf 10},  2618  (2010).

\bibitem{Oulton:2009}
R.~F. Oulton, V.~J. Sorger, T. Zentgraf, R.-M. Ma, C. Gladden, L. Dai, G.
  Bartal, and X. Zhang, {\em Plasmon lasers at deep subwavelength scale},
  Nature {\bf 461},  629  (2009).

\end{thebibliography}
\end{document}